\documentclass[a4paper]{aa}

\usepackage{natbib}
\usepackage{graphicx}
\usepackage{txfonts}
\usepackage[bookmarks=false, colorlinks=true, citecolor=blue, linkcolor=blue]{hyperref}
\usepackage{subfig}
\usepackage{blindtext}
\usepackage{upgreek}
\usepackage{pdflscape}
\usepackage{afterpage}
\usepackage{xcolor}

\newcommand{\JWST}{\textit{JWST}}

\def\micron{\hbox{\,$\upmu$m}}

\newcommand{\Msun}{\hbox{$M_{\rm \odot}$}}

\newcommand{\degree}{\ensuremath{^\circ}}
\newcommand\nodata{ ~$\cdots$~ }

\bibpunct{(}{)}{;}{a}{}{,}

\DeclareCaptionListOfFormat{subreformat}{#1#2}
\titlerunning{CO-to-H$_2$ conversion factor of outflows. Rovibrational CO emission.}
\authorrunning{Pereira-Santaella et al.}

\begin{document}

\title{The CO-to-H$_2$ conversion factor of molecular outflows. Rovibrational CO emission in NGC~3256-S resolved by \JWST\slash NIRSpec}

\author{M.~Pereira-Santaella\inst{\ref{inst1}}
\and E.~Gonz\'alez-Alfonso\inst{\ref{inst2}}
\and I.~Garc\'ia-Bernete\inst{\ref{inst3}}
\and S.~Garc\'ia-Burillo\inst{\ref{inst4}}
\and D.~Rigopoulou\inst{\ref{inst3}}}

\institute{Instituto de F\'isica Fundamental, CSIC, Calle Serrano 123, 28006 Madrid, Spain \\
\email{miguel.pereira@iff.csic.es}\label{inst1}
\and
Universidad de Alcal\'a, Departamento de F\'isica y Matem\'aticas, Campus Universitario, 28871 Alcal\'a de Henares, Madrid, Spain\label{inst2}
\and
Department of Physics, University of Oxford, Keble Road, Oxford OX1 3RH, UK\label{inst3}
\and
Observatorio Astron\'omico Nacional (OAN-IGN)-Observatorio de Madrid, Alfonso XII, 3, 28014, Madrid, Spain\label{inst4}
}

\abstract{
We analyze \JWST\slash NIRSpec observations of the CO rovibrational $v$=1--0 band at $\sim$4.67\micron\ around the dust-embedded southern active galactic nucleus (AGN) of NGC~3256 ($d$=40\,Mpc; $L_{\rm IR}$=10$^{11.6}$\,$L_\odot$).
We classify the CO $v$=1--0 spectra into three categories based on the behavior of P- and R-branches of the band: (a) both branches in absorption toward the nucleus; (b) P-R asymmetry (P-branch in emission and R-branch in absorption) along the disk of the galaxy; and (c) both branches in emission in the outflow region above and below the disk. In this paper, we focus on the outflow. The CO $v$=1--0 emission can be explained by the vibrational excitation of CO in the molecular outflow by the bright mid-IR $\sim$4.7\micron\ continuum from the AGN up to $r$$\sim$250\,pc. We model the ratios between the P($J$+2) and R($J$) transitions of the band to derive the physical properties (column density, kinetic temperature, and CO-to-H$_2$ conversion factor, $\alpha_{\rm CO}$) of the outflowing gas. We find that the $^{12}$CO $v$=1--0 emission is optically thick for $J$<4, while the $^{13}$CO $v$=1--0 emission remains optically thin.
From the P(2)\slash R(0) ratio, we identify a temperature gradient in the outflow from $>$40\,K in the central 100\,pc to $<$15\,K at 250\,pc sampling the cooling of the molecular gas in the outflow.
We used three methods to derive $\alpha_{\rm CO}$ in eight 100\,pc (0\farcs5) apertures in the outflow by fitting the P($J$+2)\slash R($J$) ratios with non-LTE models. 
We obtain low median $\alpha_{\rm CO}$ factors (0.40--0.61)$\times\frac{3.2 \times10^{-4}}{[{\rm CO}\slash {\rm H_2}]}$\,$M_\odot$ (K km\,s$^{-1}$\,pc$^2$)$^{-1}$ in the outflow regions. This implies that outflow rates and energetics might be overestimated if a ULIRG-like $\alpha_{\rm CO}$, which is 1.3--2 times larger, is assumed.
The reduced $\alpha_{\rm CO}$ can be explained if the outflowing molecular clouds are not virialized.
We also report the first extragalactic detection of a broad ($\sigma$=0.0091\micron) spectral feature at 4.645\micron\ associated with aliphatic deuterium on polycyclic aromatic hydrocarbons (D$_n$-PAH).
}

\maketitle

\section{Introduction}\label{sec:intro}

Feedback from active galactic nuclei (AGN) has been invoked to explain key properties of galaxies like the observed stellar mass function or the quenching of the star-formation (SF) in quiescent galaxies (e.g., \citealt{Somerville2015}). This SF quenching can be caused by the removal of cold gas from the host galaxy by massive outflows.
These outflows seem to be dominated by the cold molecular phase and have mass outflow rates, $\dot{M}_{\rm out}$, from a few \Msun\,yr$^{-1}$ to $>$100\,\Msun\,yr$^{-1}$ and mass loading factors, $\dot{M}_{\rm out}$ \slash star-formation rate (SFR), which can exceed 1 in extreme cases like ultra-luminous infrared galaxies (ULIRGs; \citealt{Harrison2018, Veilleux2020}).

A commonly used tracer of cold molecular outflows are rotational CO transitions in the mm and sub-mm ranges (e.g., \citealt{Combes2013, Cicone2014, GarciaBurillo2015, Pereira2018, Lutz2020, GarciaBernete2021,Lamperti2022, RamosAlmeida2022, AlonsoHerrero2018, AlonsoHerrero2019, AlonsoHerrero2023}). To transform the observed CO luminosities ($L_{\rm CO}$), into molecular masses ($M_{\rm mol}$), and the latter into outflow rates, it is necessary to apply a CO-to-H$_2$ conversion factor, $\alpha_{\rm CO}=M_{\rm mol}\slash L_{\rm CO}$, which in most cases is not constrained by the observations and a fixed value is assumed. A $\sim$1\,dex wide range of $\alpha_{\rm CO}$ between 0.3 and 2.0\,$M_\odot$ (K km\,s$^{-1}$\,pc$^2$)$^{-1}$ has been suggested for outflows from observations of multiple $[$\ion{C}{i}$]$ and rotational CO transitions in few objects (e.g., \citealt{Dasyra2016, Cicone2018, Saito2022, Ueda2022}). According to these works, possible explanations for the reduced $\alpha_{\rm CO}$ in outflows are: non-virialized molecular clouds; optically thin rotational CO emission; and\slash or CO being destroyed becoming \ion{C}{i}.
Here, we present an alternative method to estimate the $\alpha_{\rm CO}$ factor using the fundamental rovibrational CO $v$=1--0\,4.67\micron\ band that now can be observed by the NIRSpec instrument \citep{Jakobsen2022, Boker2022} on board the James Webb Space Telescope (\textit{JWST}).

The 4.67\micron\ rovibrational CO band has been previously observed in extragalactic objects from space with {\it Spitzer} and {\it AKARI} at low spectral resolution ($R$<150) which was not enough to spectrally resolve the individual transitions of the band  \citep{Spoon2004, Baba2018, Baba2022}. It has also been observed from the ground at high spectral resolution $R$$\sim$5000--10000 (to allow for the correction of the atmospheric absorption) in a few deeply obscured objects (e.g., IRAS~08572+3915, NGC~4945, and NGC~4418; \citealt{Spoon2003, Geballe2006, Shirahata2013, Onishi2021, Ohyama2023}). All these works focused on detecting the CO rovibrational band in absorption toward dusty bright IR nuclei to probe the conditions of the gas close to the central source (AGN and\slash or compact starburst). 
In this work, thanks to the sensitivity and spatial resolution of \JWST, we for the first time disentangle the rovibrational CO absorption toward the obscured AGN located in the southern nucleus of the luminous infrared (IR) merger NGC~3256 and the spatially extended CO emission around this nucleus using the integral field unit (IFU) of NIRSpec.
So far, detections of the CO band in emission have been scarce and only in Galactic sources (e.g., \citealt{GonzalezAlfonso2002, Najita2003, Rettig2005, Herczeg2011}). With \JWST, detection of the CO band in emission is likely to become common in spatially resolved objects hosting bright mid-IR sources.

\subsection{NGC~3256}

NGC~3256 is a nearby ($d$=40\,Mpc; $z$=0.00916; 190\,pc\,arcsec$^{-1}$) gas-rich advanced merger (projected nuclear separation $<$1\,kpc) with a high IR luminosity ($L_{\rm IR}$=10$^{11.6}$\,$L_\odot$) which places it in the luminous IR galaxy (LIRG) range. The southern component of NGC~3256 (NGC~3256\,S) is an edge-on disk with a position angle (PA) of $\sim$90$\degree$. It is located in front of the northern component, which is an almost face-on disk (see e.g., \citealt{Sakamoto2014}).

The northern component is powered by a bright starburst with a nuclear star-formation rate (SFR) of $\sim$15\,$M_\odot$\,yr$^{-1}$ \citep{Lira2008}.
The southern component instead hosts an extremely dust-embedded nucleus with a very deep 9.7\micron\ silicate absorption feature \citep{Pereira2010, DiazSantos2010}.
The modeling of near-IR, mid-IR, and X-ray observations of the southern nucleus suggests that it is powered by a low-luminosity ($L_{\rm 2-10\,keV}\sim10^{40}$\,erg\,s$^{-1}$) obscured AGN \citep{Ohyama2015}. However, because of the extreme dust obscuration, no high-ionization transitions neither direct X-ray emission from the AGN are detected (e.g., \citealt{Lira2002, AAH2012a, Pereira2011}).

This AGN has a collimated bipolar outflow which is perpendicular to the disk. A massive outflow ($\dot{M}_{\rm out}>$50\,$M_\odot$\,yr$^{-1}$) has been detected in the cold molecular phase, through CO, HCN, and HCO$^+$ rotational emission \citep{Sakamoto2014, Michiyama2018}, as well as in the hot molecular gas phase, through near-IR rovibrational H$_2$ transitions \citep{Emonts2014} and the ionized gas phase \citep{Leitherer2013}.
A bipolar radio-jet has been proposed as the launching mechanism of the outflow \citep{Sakamoto2014}, although it might not be powerful enough to explain the outflow mass. Alternatively, radiation pressure and an ionized wind from the AGN could drive this outflow \citep{Emonts2014}.

This paper is organized as follows: the data reduction is presented in Section~\ref{s:data}. In Section~\ref{s:analysis}, maps of the mid-IR continuum and ionized and molecular gas tracers are presented. We also extract the spectra of selected regions and fit the observed spectral features. In Section~\ref{s:co_rovib_model}, we present a model for the rovibrational CO band emission. Three methods to derive the $\alpha_{\rm CO}$ factor from the CO band emission are discussed in Section~\ref{s:conv_factor}. Finally, in Section~\ref{s:summary}, we summarize the main results of the paper.

\begin{table}
\caption{Spectral features}
\label{tbl_feature_list}
\centering
\begin{small}
\begin{tabular}{lcccccc}
\hline \hline
\\
Name & $\lambda$ & Ref. \\
& ($\mu$m) &  \\
\hline
\multicolumn{3}{c}{CO $v$=1--0 band fit}\\
\hline
H$_2$ 1--0 O(9) & 4.5755 & 2 \\
$[$\ion{K}{iii}$]$ & 4.6180 & 4\\  %
Aliphatic D$_n$-PAH & 4.645$^a$ & 5,6 \\
\ion{H}{i} 7--5 & 4.6538 & 3 \\
H$_2$ 1--1 S(10) & 4.6558 & 2 \\
\ion{H}{i} 11--6 & 4.6725 & 3 \\
$^{12}$CO $v$=1--0 & 4.6742$^b$ & 1 \\
H$_2$ 0--0 S(9) & 4.6946 & 2\\
$^{13}$CO $v$=1--0 & 4.7792$^b$ & 1 \\
C$^{18}$O $v$=1--0 & 4.7882$^b$ & 1 \\
\hline
\multicolumn{3}{c}{Additional transitions}\\
\hline
H$_2$ 0--0 S(10) & 4.4098 & 2 \\
H$_2$ 1--1 S(11) & 4.4166 & 2 \\
H$_2$ 1--1 S(9) & 4.9541 & 2 \\
H$_2$ 0--0 S(8) & 5.0531 & 2 \\
\ion{H}{i} 10--6 & 5.1287 & 3\\
\hline
\end{tabular}
\end{small}
\tablefoot{List of spectral features used during the fitting of the CO $v$=1--0 band.
$^{(a)}$ The D$_n$-PAH emission is modeled as a Gaussian profile with $\sigma$=0.0091\micron (see Section~\ref{ss:dpah} and Figure~\ref{fig_dnpah}).
$^{(b)}$ Wavelength of the P(1) transition of the rovibrational band.
}
\tablebib{(1) \citealt{Goorvitch1994}; (2) \citealt{Roueff2019}; (3) \citealt{Storey1995}; (4) \citealt{Feuchtgruber1997}; (5) \citealt{Peeters2004DPAH}; (6) This work.}
\end{table}

\section{Data reduction}\label{s:data}

\begin{figure*}
\centering
\includegraphics[width=0.8\textwidth]{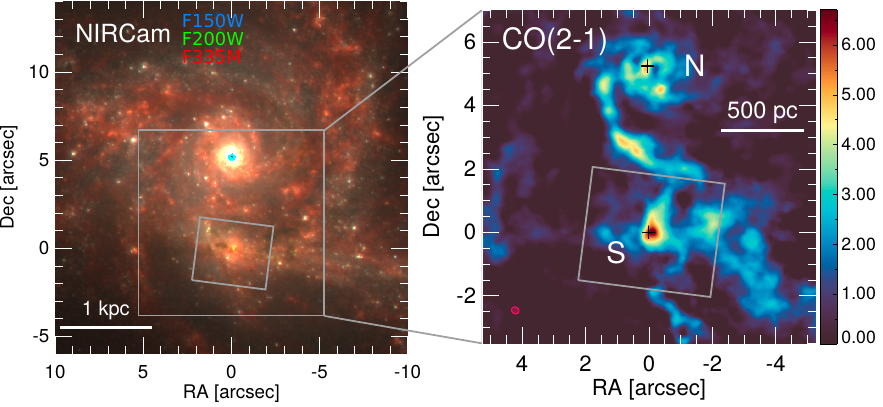}
\includegraphics[width=0.8\textwidth]{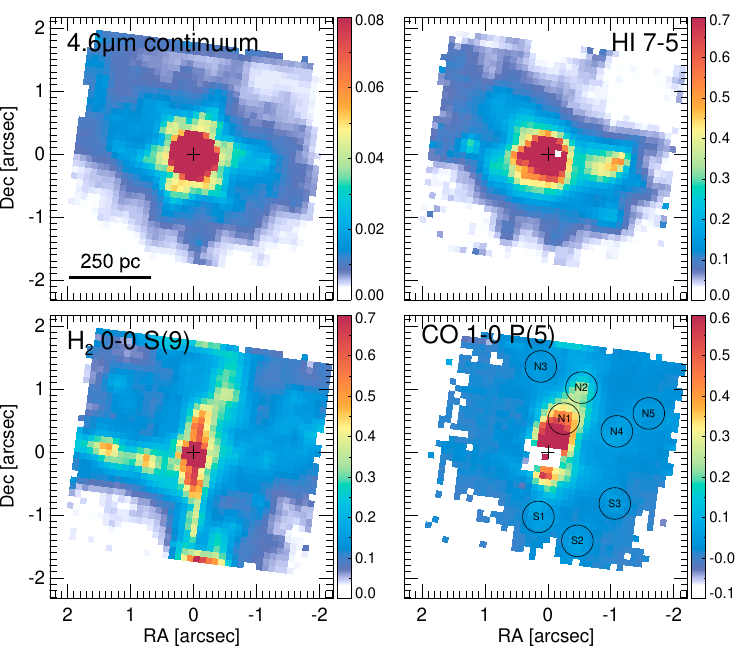}
\caption{(Top-left). Color image of NGC~3256 combining the NIRCam filters F150W (blue), F200W (green), and F335M (red). The latter contains the PAH3.3\micron\ band. The gray box marks the field of view of the ALMA CO(2--1) map (top-right panel) and the tilted gray rectangle is the NIRSpec field of view (middle and bottom row panels).
(Top-right). ALMA CO(2--1)\,230.54\,GHz zeroth moment map. The red ellipse represents the beam FWHM (0\farcs39$\times$0\farcs37). The units of the image are Jy\,km\,s$^{-1}$\,beam$^{-1}$. The crosses mark the northern and southern nuclei of NGC~3256 based on the 235\,GHz continuum.
(Middle-left). NIRSpec 4.6\micron\ continuum map (see Section~\ref{ss:spectral_maps}). The units are mJy\,pixel$^{-1}$.
(Middle-right, bottom-left, and bottom-right). Maps of the \ion{H}{i} 7--5\,4.65\micron, H$_2$ 0--0 S(9)\,4.69\micron, and CO $v$=1--0 P(5)\,4.71\micron\ transitions, respectively (see Section~\ref{ss:spectral_maps}). The maps are in units of 10$^{-16}$\,erg\,s$^{-1}$\,cm$^{-2}$\,pixel$^{-1}$. The crosses indicate the nuclear position determined from the 4.6\micron\ continuum peak. The circular apertures in the last panel are the regions used to extract the spectra of individual regions (Section~\ref{ss:sp_regions}).
\label{fig_nircam}}
\end{figure*}

\subsection{\JWST\ NIRSpec integral field spectroscopy}\label{ss:data_nirspec}

We present the high spectral resolution ($R$$\sim$1900--3600) 2.87--5.27\micron\ \JWST\slash NIRSpec integral-field spectroscopy of NGC~3256 obtained with the grating/filter pair G395H/F290LP. These NIRSpec observations are part of the Director's Discretionary Early Release Science (DD-ERS) Program \#1328 (PI: L.~Armus and A.~Evans). The data were obtained with the \textsc{NRSIR2RAPID} readout using 18 groups and a 4-point cycling dither pattern. The field of view is 3\farcs1 $\times$ 3\farcs1, with a spaxel size of 0\farcs1.

We downloaded the uncalibrated level 1 data from the \JWST\ archive\footnote{\url{https://mast.stsci.edu/portal/Mashup/Clients/Mast/Portal.html}}.
The NIRSpec IFU observations were processed using the \JWST\ Calibration pipeline (version 1.9.4) together with the calibration context 1063 from the Calibration References Data System (CRDS). First, we processed the raw data for detector-level corrections using the standard Detector1 module of the pipeline.
MSA leakage correction exposures (i.e., leakal files) are also corrected during this stage. Then, we subtracted a residual bias level by masking the slits in the ramp-fitted products (i.e., count rate files). We built a bad-pixel mask based on the leakal exposures and updated the data quality extensions of both source and leakal observations. The masked pixels are interpolated based on the neighbour spaxels (3$\times$3 box) when possible.

Then, we calibrated count rate files using the Spec2 module. This module corrects flat-fielding and applies flux calibration and world coordinate system corrections. 
We used the Spec3 module to build the two 3D spectral sub-cubes (NIRSpec detectors 1 and 2) from the fully calibrated Spec2 products. We used the drizzle weighing and spaxel size of 0\farcs1 in the cube build algorithm. 

\begin{figure*}[ht]
\centering
\includegraphics[width=17cm]{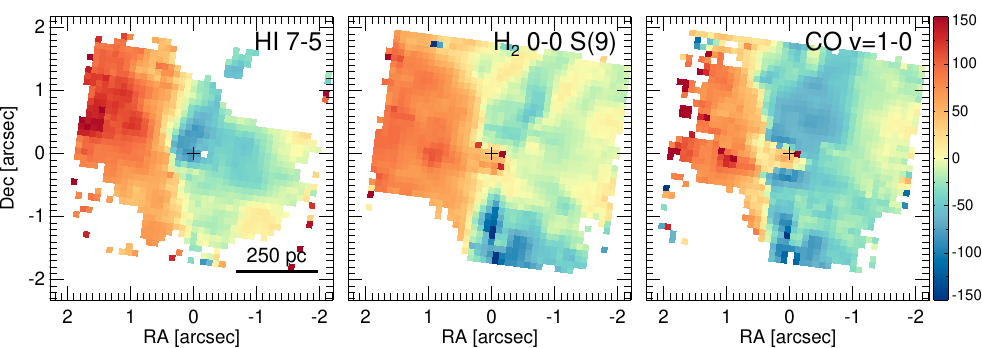}
\caption{Observed velocity fields of \ion{H}{i} 7--5\,4.65\micron\ (first panel), H$_2$ 0--0 S(9)\,4.69\micron\ (second panel), and CO $v$=1--0\,4.67\micron\ (third panel). The maps are in km\,s$^{-1}$.
\label{fig_vel}}
\end{figure*}

\subsection{NIRCam imaging}

NIRCam imaging was also available as part of the DD-ERS Program \#1328. We downloaded the fully reduced and calibrated F150W (1.50\micron), F200W (1.99\micron), and F335M (3.36\micron) images from the \JWST\ archive. We combined these three filters to produce a reference color image of the system showing the near-IR stellar morphology (F150W and F200W filters) and the more extended PAH3.3\micron\ emission included in the F335M filter (Figure~\ref{fig_nircam}).

\subsection{Ancillary ALMA data}\label{ss:data_alma}
We retrieved archival ALMA observations of the $^{12}$CO rotational lines $J$=1--0 (115.27\,GHz) and $J$=2--1 (230.54\,GHz) from programs 2018.1.00223.S (PI: K. Sakamoto) and 2015.1.00714.S (PI: K. Sliwa), respectively. We combined data from compact and extended 12-m array configurations to improve the coverage of $uv$-plane.
We also retrieved observations of the CO isotopologues $^{13}$CO $J$=2--1 (220.40\,GHz) and C$^{18}$O $J$=2--1 (219.56\,GHz), both from program 2015.1.00412.S (PI: N. Harada).

The data were processed using the ALMA reduction software CASA (v6.4.1; \citealt{McMullin2007}) following \citet{Pereira2021}. The $^{12}$CO cubes were smoothed to a common beam with a FWHM of 0\farcs39$\times$0\farcs37, which is comparable to the \JWST\slash NIRSpec resolution ($\sim$0\farcs2), using the CASA task \textsc{imsmooth}. The beam FWHM of the $^{13}$CO and C$^{18}$O cubes is 0\farcs85$\times$0\farcs81.

We calculated the zeroth moment maps of all the CO transitions following the method described in \citet{SanchezGarcia2022}. The top right panel of Figure~\ref{fig_nircam} shows the $^{12}$CO(2--1) zeroth moment map around the two nuclei of NGC~3256.

The absolute flux accuracy for Band 3 ($^{12}$CO $J$=1--0) and Band 6 ($J$=2--1 transitions) is 10\%\ and 20\%, respectively (ALMA Technical Handbook).

\section{Analysis and results}\label{s:analysis}

In this paper, we focus on the rich 4.52--4.87\micron\ spectral range observed by \JWST\slash NIRSpec where the rovibrational bands of $^{12}$CO (4.67\micron), $^{13}$CO (4.78\micron), and C$^{18}$O (4.79\micron) are located.
In addition to the CO bands, this spectral range also contains pure rotational and rovibrational H$_2$ transitions as well as hydrogen recombination lines and ionized gas tracers. Table~\ref{tbl_feature_list} shows the full list of spectral features used in this work.

To analyze the NIRSpec data cubes, we first produced spectral maps to extract the spatial information at the native NIRSpec resolution ($\sim$0\farcs2 at $\sim$4.6\micron; Sections~\ref{ss:spectral_maps} and \ref{ss:spectral_co_maps}). Then, we analyzed the spectra from selected regions of interest using larger apertures (0\farcs5 diameter; $\sim$ 100\,pc) to increase the signal-to-noise ratio (SNR; Section~\ref{ss:sp_regions}).

\begin{figure*}
\centering
\includegraphics[width=17cm]{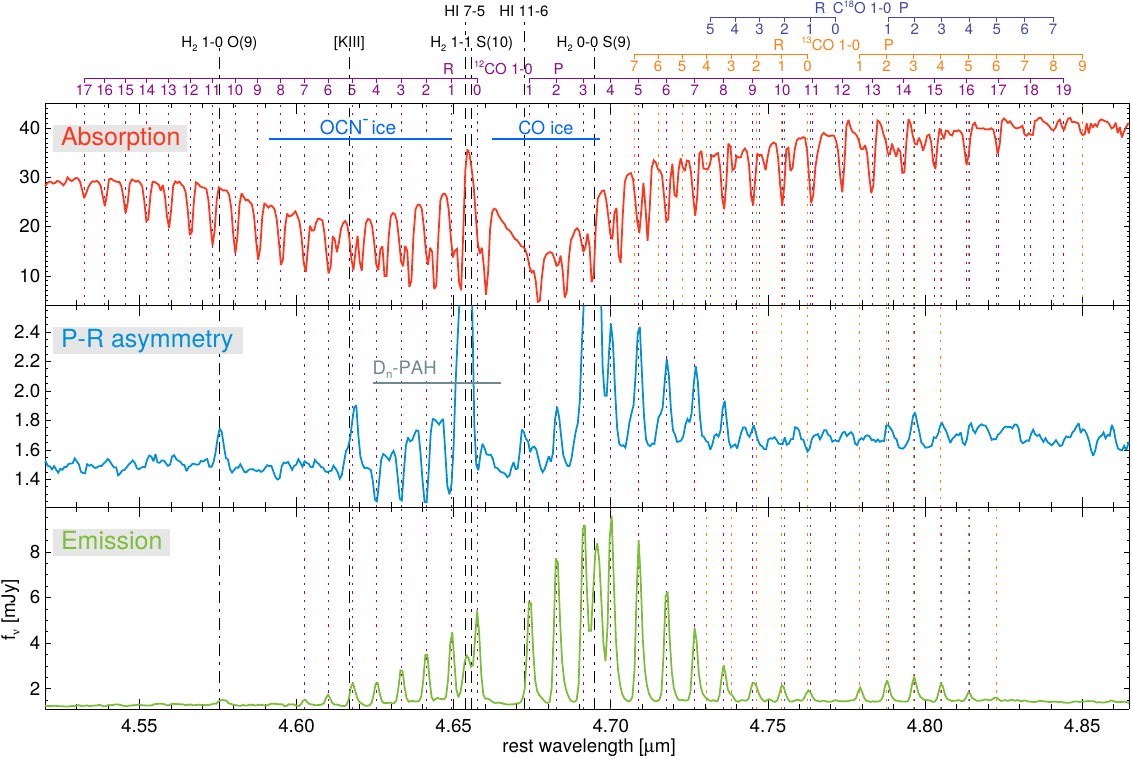}
\caption{Spectra of the CO $v$=1--0 band from the three classes defined in Section~\ref{ss:spectral_co_maps} (see Figure~\ref{fig_co_class_map}) based on the absorption\slash emission of the P- and R-branches. The top panel shows the spectrum of the nuclear region where both branches are in absorption. The middle panel is the integrated spectrum of the regions where the R-branch is in absorption and the P-branch in emission (rotating disk). The bottom panel is the integrated spectrum of areas with both branches in emission (outflow region perpendicular to the disk). The wavelength of the  P- and R-branches of the $^{12}$CO, $^{13}$CO, and C$^{18}$O $v$=1--0 bands are marked by the dashed purple, orange, and blue lines, respectively. Transitions from other atomic and molecular species are marked by the dot-dashed black lines. The D$_n$-PAH band and the OCN$^{-}$ and CO ices are indicated in the middle and top panels. The rest frame is defined using the wavelength of the higher-$J$ CO $v$=1--0 transitions detected, which for the nucleus are blueshifted with respect to the systemic velocity.
\label{fig_co_class_sp}}
\end{figure*}

\setcounter{figure}{4}
\begin{figure}
\centering
\includegraphics[width=0.35\textwidth]{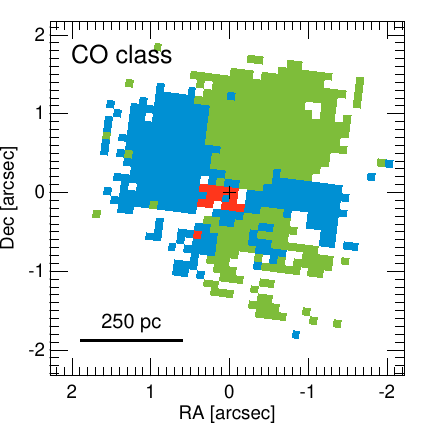}
\caption{Classification of the CO $v$=1--0 band according to the absorption\slash emission of the P- and R-branches. The red area marks regions with both branches in absorption. The blue area indicates regions where the R-branch is in absorption and the P-branch in emission. The green area are regions with both branches in emission. The spectra of these areas are presented in Figure~\ref{fig_co_class_sp}. \label{fig_co_class_map}}
\end{figure}

\subsection{Mid-IR continuum, ionized, and molecular gas maps}\label{ss:spectral_maps}

We first produced spectral maps of various tracers in the spectral range around the $^{12}$CO band (4.59--4.75\micron) to trace the morphology and kinematics of the ionized and molecular gas phases and the mid-IR continuum (Figures~\ref{fig_nircam} and \ref{fig_vel}). To do so, we extracted the spectrum of each spaxel and fitted a model including the features listed in Table~\ref{tbl_feature_list}.
For reference, Figure~\ref{fig_co_class_sp} shows the annotated spectra of three regions of NGC~3256 (see Section~\ref{ss:sp_regions} for details).

We fitted the local continuum level in the 4.59--4.75\micron\ range using a linear function. For the spectral features we used Gaussian profiles. We tied the velocity and dispersion for groups of lines tracing the same phase of the gas: warm\slash hot molecular gas (H$_2$ lines), and ionized gas (\ion{H}{i} and $[$\ion{K}{iii}$]$4.62\micron). The relative velocity of the D$_n$-PAH band (Section~\ref{ss:dpah}) is tied to the H$_2$ lines. The wavelength and width of the individual R- and P-branch CO transitions were also tied to a common line of sight velocity and velocity dispersion. In addition, the intensity of the \ion{H}{i} lines were tied according to the expected case B ratios at 10\,000\,K (7--5\slash 11--6=6.43; \citealt{Storey1995}). These two transitions are very close in wavelength (4.65 and 4.67\micron), so the differential extinction is minimal and its effect on their ratio negligible.
The intensity of the $[$\ion{K}{iii}$]$4.62\micron\ was tied to that of the \ion{H}{i} 7--5 line since the former is fainter and blended with the CO R(5) transition. We estimated an average $[$\ion{K}{iii}$]$4.62\micron\slash \ion{H}{i} 7--5\,4.65\micron\ ratio of 0.104 in this object from the average ratio measured in regions with no CO R(5) detected. To create these maps, we did not include in the model either the $^{13}$CO or the C$^{18}$O $v$=1--0 bands since they are not blended with any of these \ion{H}{i} or H$_2$ transitions.

This spectral range contains numerous overlapping spectral features, so to obtain an initial guess for the velocity and width of the lines in each spaxel, we used those derived from the H$_2$ S(8)\,5.05\micron\ and \ion{H}{i} 10--6 5.13\micron\ lines, which are isolated in a less crowded spectral range.

From these fits, we obtained the intensity maps of the 4.6\micron\ continuum, the ionized gas traced by \ion{H}{i}\,7--5\,4.65\micron\ line, the molecular gas traced by H$_2$ 0--0 S(9)\,4.69\micron, and the CO 1--0 P(5) transition, which is not strongly blended with other spectral features (see second and third rows of Figure~\ref{fig_nircam}).

\subsubsection{Southern rotating disk. Ionized, molecular, and continuum emissions}\label{ss:rotating_disk}

In addition to the compact and bright 4.6\micron\ continuum and ionized gas emission from the southern nucleus (second row of Figure~\ref{fig_nircam}), the southern disk is clearly seen extended along the East-West direction. The ionized gas is likely tracing star-forming regions in the disk, while the extended continuum is likely produced by hot dust heated by these star-forming regions.

The velocity field of the ionized gas traced by \ion{H}{i}\,7--5\,4.65\micron\ (first panel of Figure~\ref{fig_vel}) is also consistent with a distorted rotating disk due to the galaxy interaction and it is compatible with previous near-IR kinematic studies of this system \citep{Piqueras2012}. We note that the central $\sim$100\,pc velocity field is uncertain since the H$_2$ and \ion{H}{i} lines are heavily affected by the multi-component CO $v$=1--0 nuclear absorption (Section~\ref{ss:spectral_co_maps}).

The hot ($T>$1000\,K) molecular gas traced by H$_2$ 0--0 S(9)\,4.69\micron\ ($E_{\rm up}$=10261\,K; \citealt{Roueff2019}) shows a more complex morphology (third row of Figure~\ref{fig_nircam}). Some emitting regions are detected in the disk and the hot molecular gas kinematics also shows an East-West velocity gradient compatible with a distorted rotating disk (second panel of Figure~\ref{fig_vel}).
However, the brightest hot molecular emission is located perpendicular to the rotating disk and it is likely related to the massive molecular outflow (\citealt{Emonts2014}; Section~\ref{ss:molecular_outflow}).

\subsubsection{Molecular outflow. H$_2$ and CO emissions}\label{ss:molecular_outflow}

The southern AGN produces a high-velocity (${\rm v}>1000$\,km\,s$^{-1}$) bipolar collimated molecular outflow perpendicular to the rotating disk (e.g., \citealt{Sakamoto2014, Emonts2014}). The hot molecular phase of this outflow has been detected using near-IR H$_2$ transitions \citep{Emonts2014}. The H$_2$ 0--0 S(9)\,4.69\micron\ emission also traces hot molecular gas and the emission perpendicular to the disk is likely related to this outflow (see Section~\ref{ss:rotating_disk}). However, we note that due to the small field of view of NIRSpec, the brightest hot H$_2$ clumps of the outflow, which also have the highest velocity dispersion ($\sigma\sim$190\,km\,s$^{-1}$; \citealt{Emonts2014, Piqueras2012}), are not covered by these observations.

The cold molecular phase of the outflow is traced by the sub-mm\slash mm rotational CO transitions \citep{Sakamoto2014}. The top row of Figure~\ref{fig_nircam} shows the morphology of the CO(2--1)\,230.54\,GHz transition. Although some CO(2--1) emission is detected in the southern disk, most of the emission around this nucleus is instead extended perpendicular to the disk up to $\sim$400\,pc away. Thanks to the high angular resolution of the ALMA data, it is clear that this emission is not part of the spiral arms of the northern galaxy and, therefore, it is likely located above and below the southern nucleus and its origin likely connected to the molecular outflow. Consequently, we consider the emission from these regions as part of the outflow.
Moreover, the outflow direction is close to the plane of the sky ($i\sim80\deg$; \citealt{Sakamoto2014}), so even if the observed line of sight velocity is not particularly high (${\rm v}$<150\,km\,s$^{-1}$), the inclination corrected velocity can be much higher (see \citealt{Sakamoto2014}).

The bottom row of Figure~\ref{fig_nircam} presents the CO $v$=1--0 rovibrational emission map as traced by the P(5) transition. It is similar to the CO(2--1)\,230.54\,GHz emission, and it is relatively brighter in the  outflow direction compared to the disk emission (see Section~\ref{ss:spectral_co_maps}). In Figure~\ref{fig_pv_co}, we compare the position-velocity (PV) plot along the outflow axis for the CO(2--1) and P(5) transitions. Excluding the strong nuclear absorption affecting the P(5) transition (Section~\ref{ss:spectral_co_maps}), the two PV plots show a similar morphology. In particular, the higher-velocity blue-shifted ($-$100--$-$200\,km\,s$^{-1}$) outflow emission between $-$1\farcs1 and $-$0\farcs3 is detected in both transitions.
Thus, based on the similar morphology and kinematics, we conclude that the rovibrational CO emission is related to the cold molecular outflow identified in the ALMA data.

\setcounter{figure}{3}
\begin{figure}
\centering
\includegraphics[width=0.348\textwidth]{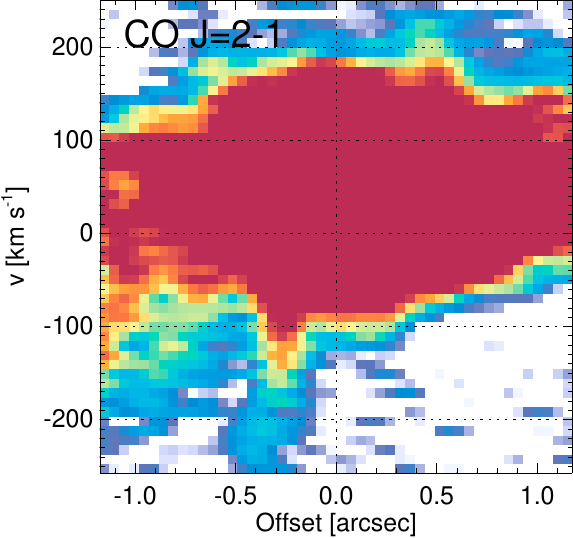}
\includegraphics[width=0.35\textwidth]{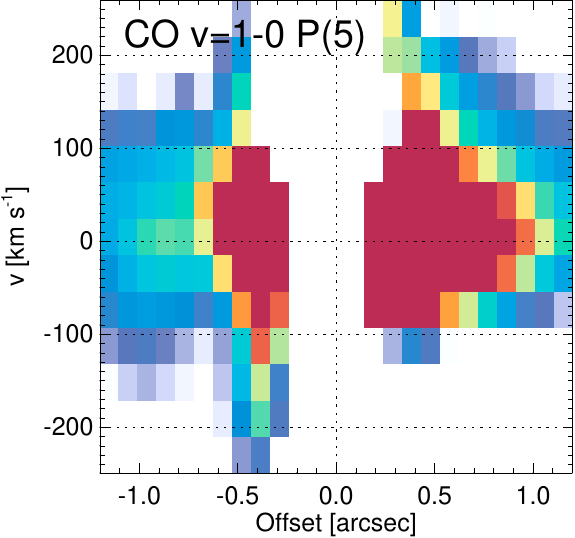}
\caption{ Position-velocity cuts taken along the outflow axis (PA=0\degree) of the CO(2--1)\,230.54\,GHz (top) and the CO \hbox{$v$=1--0} P(5)\,4.709\micron\ (bottom) transitions. For the CO(2--1), we used the spatially un-smoothed cube with a beam FWHM of 0\farcs23$\times$0\farcs20. The nuclear P(5) absorption around offset$\sim$0\arcsec\ is masked in the bottom panel. Positive offsets are north of the nucleus.
\label{fig_pv_co}}
\end{figure}
\setcounter{figure}{5}

\subsection{Origin of the rovibrational CO $v$=1--0 band}\label{ss:spectral_co_maps}

The ratio between the P- and R-branch transitions of the CO $v$=1--0 band can be used to investigate the physical conditions of the molecular gas as traced by CO.

\begin{figure}
\centering
\includegraphics[width=0.45\textwidth]{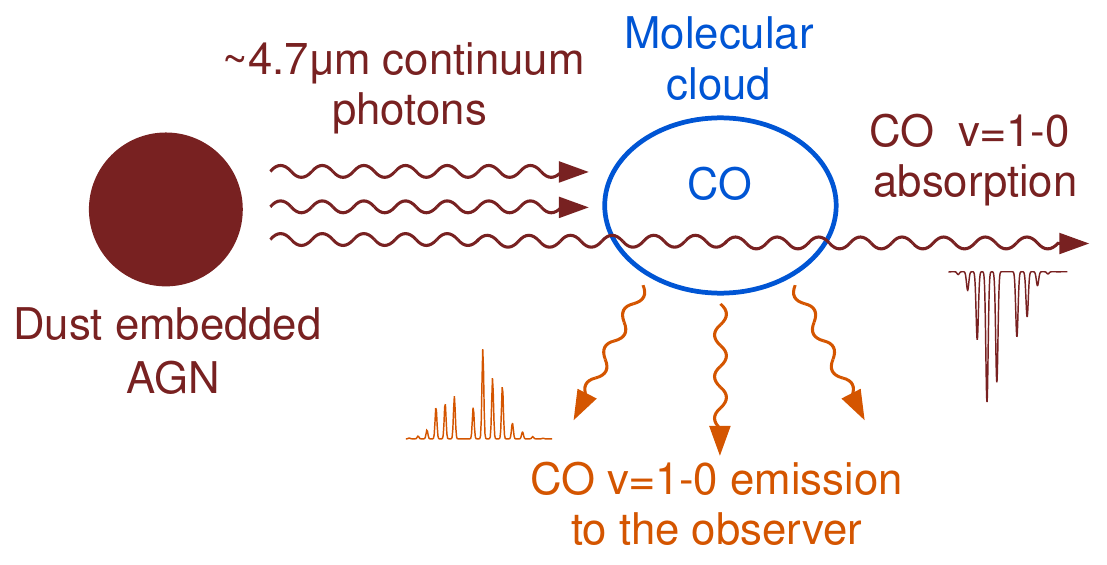}
\caption{Sketch showing how the CO $v$=1--0 band emission is produced in the outflow region of NGC~3256\,S.
The $\sim$4.7\micron\ photons from the dust around the AGN (brown lines) illuminate the molecular clouds (blue ellipse). These photons are absorbed by CO molecules which are excited to the $v$=1 levels and decay through the P- and R-branches (orange lines). The CO $v$=1--0 absorption is only observed in lines of sight that intersect both the $\sim$4.7\micron\ continuum source and the molecular clouds.
\label{fig_sketch}}
\end{figure}

The R(2)\,4.641\micron\ and P(5)\,4.709\micron\ transitions are not blended with other features so they are ideal to compare the P- and R-branches. We fitted a Gaussian profile to these two lines with tied velocity and velocity dispersion. We leave the peak of the Gaussian free allowing for both absorption and emission profiles.
The R(2) and P(5) transitions are detected in emission and absorption in different regions. According to the emission or absorption profiles, we classify the spaxels in three groups: (a) both transitions in absorption; (b) R(2) in absorption and P(5) in emission (P-R asymmetry); and (c) both transitions in emission.
The integrated spectra of these regions are shown in Figure~\ref{fig_co_class_sp} and the spatial classification in Figure~\ref{fig_co_class_map}.
From this map it is clear that (a) spaxels correspond to the nucleus where the strong 4.6\micron\ continuum favors the detection of the CO $v$=1--0 band in absorption. Also, we note that the nuclear CO $v$=1--0 spectrum has two velocity components: a low-velocity colder component only detected in the low-$J$ transitions (up to $J$$\sim$6); and a blue-shifted (v=160\,km\,s$^{-1}$) possibly outflowing component detected up to $J$=19.
The (b) spaxels showing the P-R asymmetry are mostly located along the disk of the galaxy. This P-R asymmetry has  been detected in Galactic star-forming regions (e.g., \citealt{GonzalezAlfonso2002, GonzalezAlfonso1998}), which is consistent with the spatial location of these regions in NGC~3256.
Finally, the (c) spaxels with both P- and R-branches in emission are detected perpendicular to the disk in the molecular outflow region. The CO $v$=1--0 band can be radiatively excited by $\sim$4.7\micron\ photons.
Therefore, it is possible to explain this CO $v$=1--0 emission if the mid-IR radiation from the AGN is illuminating the cold molecular outflowing gas (see Figure~\ref{fig_sketch}).
This scenario and the method to derive the cold molecular gas properties from the  CO $v$=1--0 band are further explored in Sections~\ref{s:co_rovib_model} and \ref{s:conv_factor}.

\subsection{Spectra from regions with CO $v$=1--0 in emission}\label{ss:sp_regions}

We extracted the spectra from eight circular apertures (0\farcs5 diameter) in the outflow region where both the P- and R-branches are detected in emission (green area in Figure~\ref{fig_co_class_map}; see also Section~\ref{ss:spectral_co_maps}). The location of these apertures is plotted in the bottom-right panel of Figure~\ref{fig_nircam}. These apertures follow the outflow axis (N1, N2, and S1) and few clumps seen in the CO $v$=1--0 map (Figure~\ref{fig_nircam} bottom right panel; N3, N4, N5, S2, and S3).

Since the nuclear emission is compact and relatively bright (Figure~\ref{fig_nircam}), we need to subtract its contribution to the spectrum of these apertures. We calculated this contribution at each wavelength using point-spread functions (PSFs) created with WebbPSF (v1.1.1; \citealt{Perrin2014}). These PSFs were rotated and convolved with a Gaussian kernel (FWHM=0\farcs130$\times$0\farcs077) to match the orientation and FWHM of the southern nucleus of NGC~3256. Using these PSFs, we estimated the scaling factor between the emission at the aperture position and that from a fiducial aperture (0\farcs4 diameter) centered on the PSF. Then, using the fiducial aperture, we measured the NGC~3256\,S nuclear emission and applied the scaling factor to subtract its contribution to the aperture spectrum.

We fitted the spectral features in the 4.52--4.87\micron\ range following a procedure similar to that described in Section~\ref{ss:spectral_maps}. However, because of the higher SNR of these spectra, compared to the individual spaxel spectra, we used a more detailed model which includes additional spectral features and components. We describe the main elements of this more detailed model in Sections~\ref{ss:baseline} through \ref{ss:12co_band}.

\subsubsection{Continuum baseline}\label{ss:baseline}

The shape of the continuum around the CO $v$=1--0 rovibrational band can be affected by absorption bands produced by various ice species (OCN$^-$ 4.62\micron, $^{12}$CO 4.67\micron, and $^{13}$CO 4.78\micron; e.g., \citealt{Gibb2004, McClure2023}).

\begin{figure}
\centering
\includegraphics[width=0.42\textwidth]{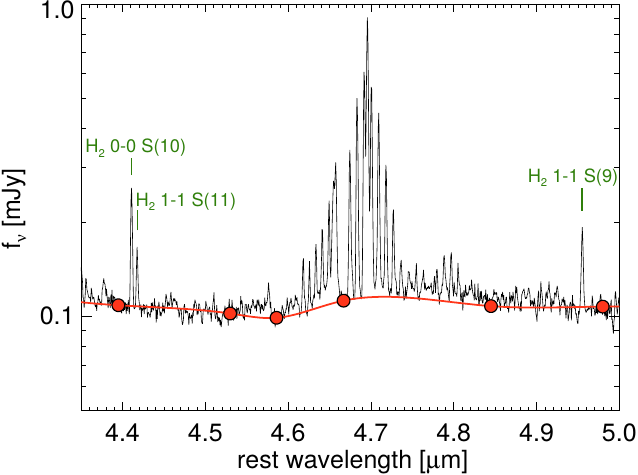}
\caption{Spectrum from region N3 (black line) and the spline-interpolated continuum baseline (red line). The pivot points are indicated by the red circles (see Section~\ref{ss:baseline}). The wavelengths of the three H$_2$ transitions not presented in Figures~\ref{fig_co_class_sp} and \ref{fig_fit_model} are indicated in green (see also Table~\ref{tbl_feature_list}).
\label{fig_cont_spline}}
\end{figure}

To take into account these effects on the continuum shape, we defined a spline-interpolated baseline. The pivot points (4.395, 4.530, 4.586, 4.667, 4.845, and 4.980\micron) are selected to track the continuum level avoiding known emission features. This baseline reproduces well the underlying continuum and ice absorptions (e.g., Figure~\ref{fig_cont_spline}) and was subtracted from the spectra before fitting the emission lines.

\subsubsection{Optically thin $^{13}$CO and C$^{18}$O $v$=1--0 bands}\label{ss:13cothin}

The R-branch of the $^{13}$CO and C$^{18}$O $v$=1--0 bands overlaps with the P-branch of the $^{12}$CO band (see Figure~\ref{fig_co_class_sp}).
Although, the $^{13}$CO and C$^{18}$O bands are weaker than the band produced by the main isotopologue $^{12}$CO, for P-branch transitions with $J>$8, the intensities of the overlapping R-branch transitions from the isotopologues are comparable or brighter than the $^{12}$CO lines. To disentangle the bands from the three species, we used Gaussian profiles to fit the P-branch and the R(0) transition of $^{13}$CO, which are not affected by $^{12}$CO band. The widths and velocities of these profiles were tied and we fitted them independently of the kinematics obtained for the $^{12}$CO band.
The $^{13}$CO P(2)/R(0) ratio is consistent with the optically thin limit, $\sim$1.96 (see Section~\ref{ss:model_p2r}), within $\pm$1\,$\sigma$ for all the regions (except for S3 where the $^{13}$CO band spectral range is noisier).
Therefore, it is reasonable to assume that the $^{13}$CO (and also the C$^{18}$O) emission is optically thin. Under this assumption, it is possible to determine the flux of the R-branch using the measured P-branch fluxes (Section~\ref{ss:model_p2r}). We estimate the intensity of the C$^{18}$O band assuming that both $^{13}$CO and C$^{18}$O share the same excitation temperatures and have abundance ratio of $\sim$7.4. We derived this abundance ratio from the $^{13}$CO(2--1)\slash C$^{18}$O(2--1) ratio measured at the southern nucleus using a circular aperture ($r$=1\arcsec). This abundance ratio agrees with that obtained by \citet{Harada2018}, 5.6--6.6, for the same region using a slightly larger aperture.

The model for the $^{13}$CO and C$^{18}$O band emission is also subtracted from the spectra of the eight circular apertures before fitting the $^{12}$CO band. 
Under the optically thin limit, the fluxes predicted for the R-branch lines based on the P-branch observations, agree with the observed spectra for all the regions including region S3 (Figure~\ref{fig_fit_model}).

\subsubsection{Molecular hydrogen and recombination lines}\label{ss:h2_hi_lines}

There are two H$_2$ transitions (0--0 S(9)~4.69\micron\ and 1--1 S(10)~4.66\micron) and two \ion{H}{I} recombination lines (7--5~4.65\micron\ and 11--6~4.67\micron) that lie close to the center of the $^{12}$CO band. The fluxes of the S(9)~4.69\micron\ and 7--5~4.65\micron\ transitions are similar to the CO band fluxes, while the fluxes of the remaining two transitions are fainter.

To help constraining the contribution of all these lines, we make use of additional H$_2$ and \ion{H}{I} recombination lines in the NIRSpec range (see Table~\ref{tbl_feature_list}).

We determined the profile of the H$_2$ lines by fitting a Gaussian (or two Gaussians when a single Gaussian cannot reproduce the observed profile) to the 0--0 S(8)~5.05\micron\ profile. For the 0--0 S(9)~4.69\micron\ transition, we used this profile fixing its wavelength (velocity) and width (velocity dispersion) and allowing only the integrated flux to vary during the fit.
The fainter 1--1 S(10)~4.66\micron, is very close in wavelength to the CO~R(0) and 7--5 transitions so its flux is not well constrained. Therefore, we estimated its intensity assuming an excitation temperature equal to that derived from the 1--1 S(11)~4.42\micron\ and 1--1 S(9)~4.95\micron\ transitions and an ortho-to-para ratio of 3.

We fixed the velocity and width of the \ion{H}{I} lines taking the \ion{H}{I} 10--6\,5.13\micron\ profile as reference. The integrated flux of the 7--5 transition is fitted while for the 11--6 transition we assume the case B ratio (see Section~\ref{ss:spectral_maps}).

\subsubsection{The 4.65\micron\ D$_n$-PAH profile}\label{ss:dpah}

The aliphatic C-D stretch in deuterated PAH (D$_n$-PAH) produces a broad band at $\sim$4.65\micron. This band has been detected in Galactic sources and is predicted by quantum chemical computations (e.g., \citealt{Peeters2004DPAH, Onaka2014, Doney2016, Buragohain2020, Yang2023}). An accurate profile of this feature is not yet available due to the limited SNR and\slash or spectral resolution of these observations. Therefore, we determined the D$_n$-PAH profile from the observations of NGC~3256-S.

\begin{figure}
\centering
\includegraphics[width=0.42\textwidth]{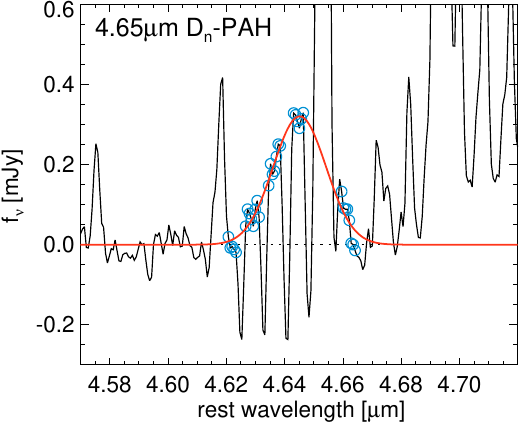}
\caption{Gaussian fit to the 4.65\micron\ D$_n$-PAH profile (red line). We masked the CO R-branch absorptions in the spectrum of the P-R asymmetry region (solid black line) so that only the blue circles are considered for this fit (see Section~\ref{ss:dpah}).
\label{fig_dnpah}}
\end{figure}

We used the spectrum from the P-R asymmetry region (middle panel of Figure~\ref{fig_co_class_sp}) where the low-$J$ R-branch transitions of the CO $v$=1--0 band are detected in absorption over the 4.65\micron\ D$_n$-PAH feature. We fitted a Gaussian to the continuum subtracted spectrum between 4.62 and 4.67\micron\ after masking the CO absorptions and the \ion{H}{I} 7--5 and H$_2$ 1--1 S(10) transitions (Figure~\ref{fig_dnpah}). The redshift of this spectrum was determined using the H$_2$ 0--0 S(8) line as reference.
The central wavelength of the best-fit Gaussian is 4.6452 $\pm$ 0.0004 $\pm$ 0.0014\micron, where the uncertainties are statistical and due to the redshift, respectively. The width of this Gaussian profile is $\sigma$=0.0091 $\pm$ 0.0003\micron. The D$_n$-PAH band has been detected in Galactic sources using NIRSpec observations \citep{Boersma2023}. The central wavelength and width we obtained are compatible with the Galactic profile, although we note that they report a SNR of the band lower than in the spectrum analyzed here (5.25 vs. 11).

\subsubsection{$^{12}$CO $v$=1--0 band}\label{ss:12co_band}

After subtracting the continuum baseline (Section~\ref{ss:baseline}) and the $^{13}$CO and C$^{18}$O bands (Section~\ref{ss:13cothin}), we fitted the $^{12}$CO band together with the H$_2$ and \ion{H}{I} transitions (Section~\ref{ss:h2_hi_lines}) and the 4.65\micron\ D$_n$-PAH band (Section~\ref{ss:dpah}) considering all the constraints described before.

We first attempted the fit using a single Gaussian profile for each CO transition. Two regions (N1 and N2) show a second broader kinematic component in the residuals. For these regions, we repeated the fit using two Gaussians for the CO band. We note that the fluxes of the broad component are not well constrained for the R(2) and R(3) transitions which are close to the D$_n$-PAH band and the \ion{H}{I} 7--5 transition. Thus, we did not attempt to disentangle the fluxes of the narrow and broad components for these regions and only considered the integrated flux of the line.

\begin{figure*}
\centering
\includegraphics[width=0.495\textwidth]{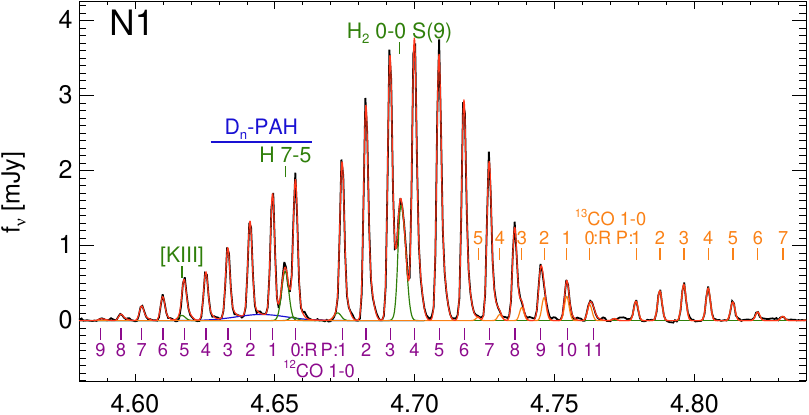}
\includegraphics[width=0.495\textwidth]{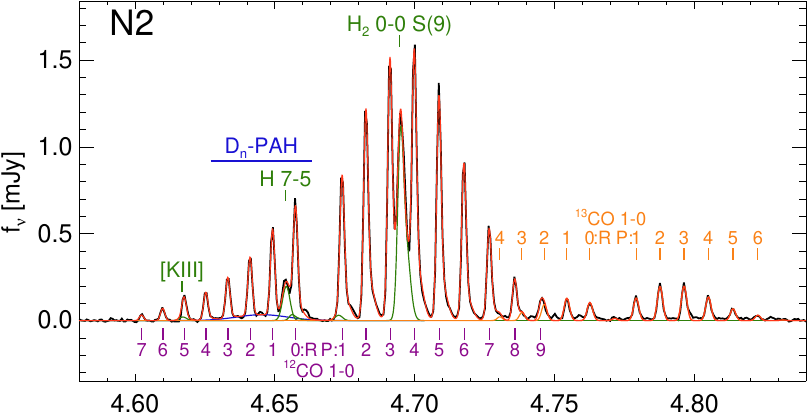}
\includegraphics[width=0.495\textwidth]{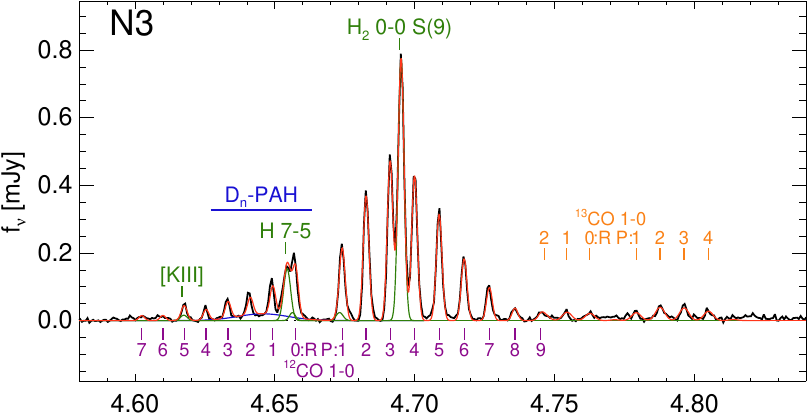}
\includegraphics[width=0.495\textwidth]{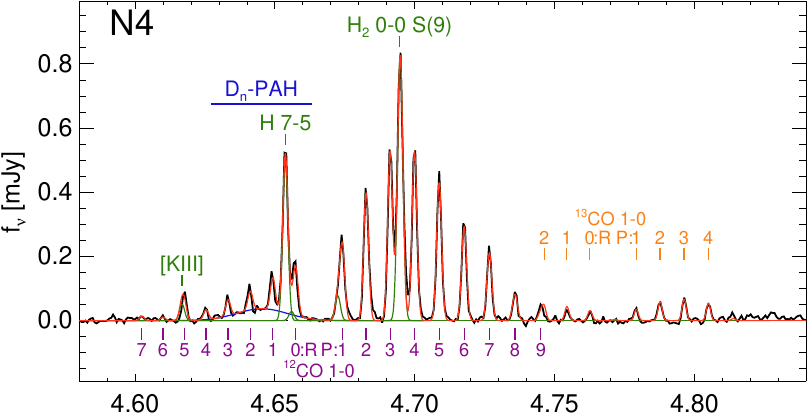}
\includegraphics[width=0.495\textwidth]{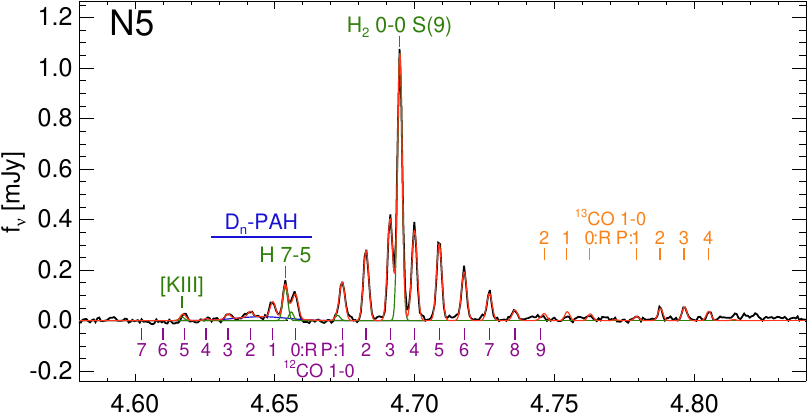}
\includegraphics[width=0.495\textwidth]{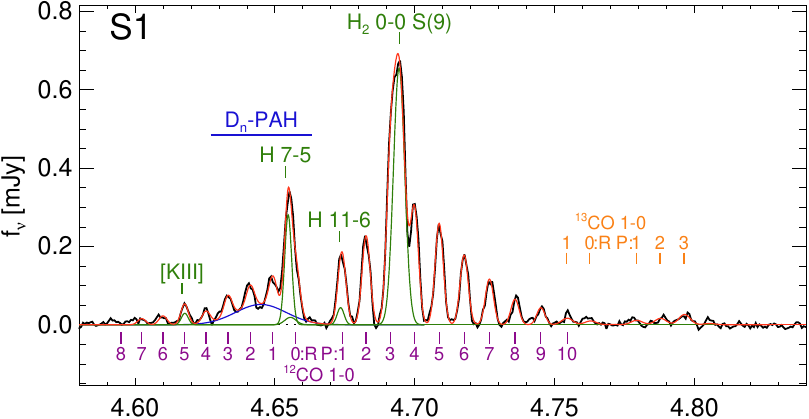}
\includegraphics[width=0.495\textwidth]{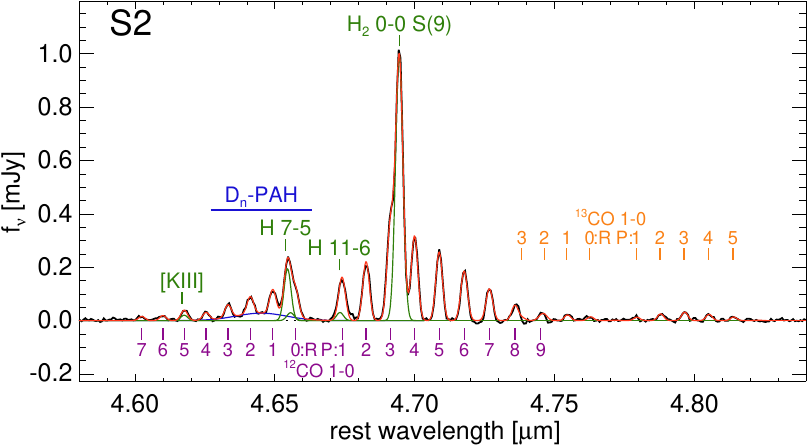}
\includegraphics[width=0.495\textwidth]{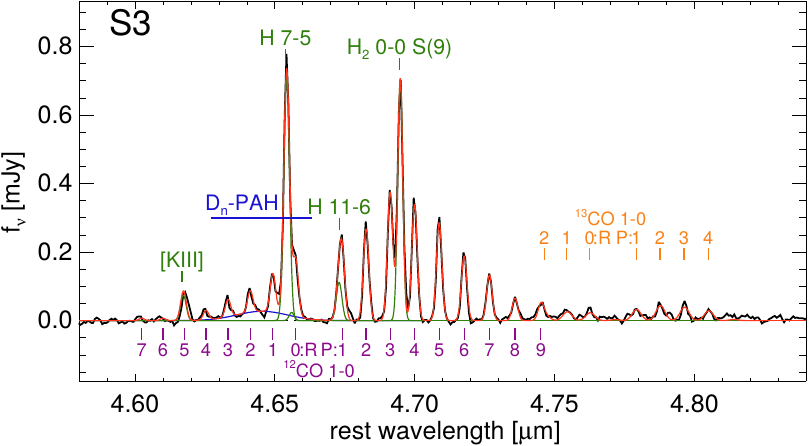}
\caption{Continuum subtracted spectra (black line) of the selected regions (see Section~\ref{ss:sp_regions} and bottom right panel of Figure~\ref{fig_nircam}) where the CO $v$=1--0 band is detected in emission. The red line is the best-fit model. The $^{13}$CO and C$^{18}$O emission lines are fitted and subtracted (see Section~\ref{ss:13cothin}) before fitting the remaining spectral features.
The green, blue, and orange lines, which are also included in the best-fit, represent the atomic and molecular transitions, D$_n$-PAH feature, and $^{13}$CO $v$=1--0 band, respectively. The wavelengths of the $^{12}$CO $v$=1--0 transitions are indicated in purple. All the spectral features present in the best-fit model are listed in Table~\ref{tbl_feature_list}.
\label{fig_fit_model}}
\end{figure*}

The best-fit model for each region are shown in Figure~\ref{fig_fit_model}. From these fits, we obtained the fluxes of the P- and R-branch transitions (Table~\ref{tbl_line_fluxes}) which we used to calculate the ratios discussed in Sections~\ref{s:co_rovib_model} and \ref{s:conv_factor}.

\begin{table*}
\caption{Fluxes and velocity dispersion of the $^{12}$CO and $^{13}$CO $v$=1--0 bands}
\label{tbl_line_fluxes}
\centering
\begin{small}
\begin{tabular}{lccccccccccccc}
\hline \hline
\\
Transition & Wavelength & \multicolumn{8}{c}{Region} \\[0.5ex]
\cline{3-10} \\[-2ex]
& \micron & N1 & N2 & N3 & N4 & N5 & S1 & S2 & S3 \\[0.5ex]
\hline
$^{12}$CO $v$=1--0\\
$\sigma_1 \slash$(km\,s$^{-1}$)~\tablefootmark{a} & & 34 $\pm$ 1 & 36 $\pm$ 1  & 55 $\pm$ 2 & 45 $\pm$ 2 & 51 $\pm$ 3 & 86 $\pm$ 3 & 67 $\pm$ 3 & 52 $\pm$ 2\\
$\sigma_2 \slash$(km\,s$^{-1}$)~\tablefootmark{a} & & 78 $\pm$ 3 & 105 $\pm$ 4 & \nodata & \nodata & \nodata & \nodata & \nodata & \nodata  \\
\hline
R(9) & 4.5876 &  0.80 $\pm$  0.64 & \nodata & \nodata & \nodata & \nodata & \nodata & \nodata & \nodata \\
R(8) & 4.5950 &  2.68 $\pm$  0.64 & \nodata & \nodata & \nodata & \nodata & \nodata & \nodata & \nodata \\
R(7) & 4.6024 &  5.98 $\pm$  0.64 &  0.93 $\pm$  0.28 &  0.52 $\pm$  0.10 &  0.41 $\pm$  0.11 &  0.00 $\pm$  0.09 &  0.78 $\pm$  0.14 &  0.63 $\pm$  0.10 &  0.23 $\pm$  0.11 \\
R(6) & 4.6100 &  9.53 $\pm$  0.64 &  2.05 $\pm$  0.28 &  0.54 $\pm$  0.10 &  0.34 $\pm$  0.11 &  0.00 $\pm$  0.08 &  1.14 $\pm$  0.14 &  0.76 $\pm$  0.10 &  0.24 $\pm$  0.11 \\
R(5) & 4.6177 &  15.1 $\pm$   0.6 &  3.51 $\pm$  0.28 &  1.01 $\pm$  0.10 &  1.22 $\pm$  0.21 &  0.50 $\pm$  0.10 &  1.18 $\pm$  0.14 &  0.76 $\pm$  0.10 &  0.44 $\pm$  0.26 \\
R(4) & 4.6254 &  19.0 $\pm$   0.6 &  4.79 $\pm$  0.28 &  1.11 $\pm$  0.10 &  0.94 $\pm$  0.11 &  0.21 $\pm$  0.10 &  1.52 $\pm$  0.14 &  1.16 $\pm$  0.10 &  0.93 $\pm$  0.15 \\
R(3) & 4.6333 &  27.7 $\pm$   0.7 &  7.06 $\pm$  0.28 &  1.72 $\pm$  0.11 &  1.54 $\pm$  0.12 &  0.68 $\pm$  0.11 &  2.64 $\pm$  0.17 &  1.94 $\pm$  0.10 &  1.73 $\pm$  0.11 \\
R(2) & 4.6412 &  36.8 $\pm$   0.7 &  10.0 $\pm$   0.3 &  1.80 $\pm$  0.19 &  1.78 $\pm$  0.20 &  0.76 $\pm$  0.12 &  2.69 $\pm$  0.20 &  2.64 $\pm$  0.12 &  2.13 $\pm$  0.12 \\
R(1) & 4.6493 &  46.6 $\pm$   1.1 &  14.8 $\pm$   0.4 &  2.97 $\pm$  0.27 &  3.07 $\pm$  0.21 &  2.04 $\pm$  0.12 &  3.91 $\pm$  0.31 &  3.74 $\pm$  0.12 &  3.82 $\pm$  0.22 \\
R(0) & 4.6575 &  52.5 $\pm$   1.3 &  18.8 $\pm$   0.5 &  4.77 $\pm$  0.25 &  4.45 $\pm$  0.26 &  3.07 $\pm$  0.12 &  6.28 $\pm$  0.29 &  3.86 $\pm$  0.11 &  5.53 $\pm$  0.23 \\
P(1) & 4.6742 &  60.7 $\pm$   1.1 &  24.8 $\pm$   0.6 &  6.91 $\pm$  0.16 &  6.86 $\pm$  0.14 &  4.91 $\pm$  0.12 &  7.25 $\pm$  0.18 &  5.60 $\pm$  0.15 &  5.69 $\pm$  0.13 \\
P(2) & 4.6826 &  85.2 $\pm$   1.6 &  37.2 $\pm$   0.8 &  13.0 $\pm$   0.2 &  12.1 $\pm$   0.2 &  9.18 $\pm$  0.12 &  11.3 $\pm$   0.3 &  8.92 $\pm$  0.16 &  9.01 $\pm$  0.16 \\
P(3) & 4.6912 & 105.6 $\pm$   2.1 &  46.0 $\pm$   0.9 &  16.5 $\pm$   0.2 &  16.0 $\pm$   0.3 &  13.4 $\pm$   0.3 &  15.7 $\pm$   0.5 &  13.7 $\pm$   0.3 &  12.6 $\pm$   0.3 \\
P(4) & 4.6999 & 112.1 $\pm$   2.5 &  45.4 $\pm$   1.2 &  15.0 $\pm$   0.2 &  16.0 $\pm$   0.4 &  11.8 $\pm$   0.4 &  14.3 $\pm$   0.5 &  12.8 $\pm$   0.2 &  11.4 $\pm$   0.3 \\
P(5) & 4.7088 & 105.4 $\pm$   2.1 &  39.0 $\pm$   0.8 &  11.0 $\pm$   0.1 &  13.0 $\pm$   0.1 &  9.98 $\pm$  0.15 &  12.8 $\pm$   0.2 &  10.3 $\pm$   0.1 &  9.55 $\pm$  0.12 \\
P(6) & 4.7177 &  86.3 $\pm$   1.7 &  27.0 $\pm$   0.6 &  6.36 $\pm$  0.12 &  8.84 $\pm$  0.14 &  6.31 $\pm$  0.15 &  8.75 $\pm$  0.18 &  7.45 $\pm$  0.11 &  6.28 $\pm$  0.11 \\
P(7) & 4.7267 &  62.0 $\pm$   1.3 &  15.7 $\pm$   0.3 &  3.40 $\pm$  0.13 &  6.34 $\pm$  0.12 &  3.50 $\pm$  0.11 &  5.72 $\pm$  0.19 &  4.77 $\pm$  0.10 &  4.44 $\pm$  0.11 \\
P(8) & 4.7359 &  35.8 $\pm$   0.7 &  6.86 $\pm$  0.27 &  1.27 $\pm$  0.10 &  2.46 $\pm$  0.11 &  1.35 $\pm$  0.10 &  3.19 $\pm$  0.19 &  2.30 $\pm$  0.13 &  2.17 $\pm$  0.10 \\
P(9) & 4.7451 &  17.4 $\pm$   0.6 &  2.20 $\pm$  0.26 &  0.35 $\pm$  0.10 &  0.84 $\pm$  0.17 &  0.25 $\pm$  0.15 &  2.12 $\pm$  0.14 &  0.66 $\pm$  0.13 &  1.31 $\pm$  0.10 \\
P(10) & 4.7545 &  6.32 $\pm$  0.61 & \nodata & \nodata & \nodata & \nodata & \nodata & \nodata & \nodata \\
P(11) & 4.7640 &  3.73 $\pm$  0.61 & \nodata & \nodata & \nodata & \nodata & \nodata & \nodata & \nodata \\
\hline
$^{13}$CO $v$=1--0\\
$\sigma \slash$(km\,s$^{-1}$)~\tablefootmark{a} & & 39 $\pm$ 6 & 50 $\pm$ 4 & 100 $\pm$ 20 & 30 $\pm$ 5 & 45 $\pm$ 6 & 130 $\pm$ 20 & 63 $\pm$ 5 & 90 $\pm$ 15 \\
\hline
R(0) & 4.7626 & 7.66 $\pm$ 0.62~\tablefootmark{b} & 3.26 $\pm$ 0.16 & 1.31 $\pm$ 0.16 & 0.62 $\pm$ 0.08 & 0.62 $\pm$ 0.10 & 1.25 $\pm$ 0.21~\tablefootmark{b} & 0.50 $\pm$ 0.09 & 1.63 $\pm$ 0.20 \\
P(1) & 4.7792 & 6.61 $\pm$ 0.38 & 4.07 $\pm$ 0.15 & 1.27 $\pm$ 0.15 & 0.85 $\pm$ 0.08 & 0.48 $\pm$ 0.10 & 0.75 $\pm$ 0.13 & 0.47 $\pm$ 0.06 & 1.69 $\pm$ 0.14 \\
P(2) & 4.7877 & 9.34 $\pm$ 0.42 & 5.66 $\pm$ 0.18 & 1.97 $\pm$ 0.14 & 1.32 $\pm$ 0.09 & 1.34 $\pm$ 0.13 & 0.99 $\pm$ 0.14 & 0.86 $\pm$ 0.06 & 2.06 $\pm$ 0.17 \\
P(3) & 4.7963 & 11.1 $\pm$ 0.5 & 5.49 $\pm$ 0.19 & 1.76 $\pm$ 0.13 & 1.44 $\pm$ 0.10 & 1.39 $\pm$ 0.14 & 1.61 $\pm$ 0.16 & 1.05 $\pm$ 0.07 & 1.77 $\pm$ 0.17 \\
P(4) & 4.8050 & 9.54 $\pm$ 0.43 & 3.53 $\pm$ 0.16 & 1.26 $\pm$ 0.14 & 1.13 $\pm$ 0.09 & 0.89 $\pm$ 0.11 & \nodata & 0.71 $\pm$ 0.07 & 1.26 $\pm$ 0.16 \\
P(5) & 4.8138 & 5.04 $\pm$ 0.37 & 1.64 $\pm$ 0.14 & \nodata & \nodata & \nodata & \nodata & 0.39 $\pm$ 0.06 & \nodata \\
P(6) & 4.8227 & 2.22 $\pm$ 0.35 & 0.74 $\pm$ 0.13 & \nodata & \nodata & \nodata & \nodata & \nodata & \nodata \\
P(7) & 4.8317 & 1.12 $\pm$ 0.34 & \nodata & \nodata & \nodata & \nodata & \nodata & \nodata & \nodata \\
\hline
\end{tabular}
\end{small}
\tablefoot{The fluxes are in 10$^{-17}$\,erg\,cm$^{-2}$\,s$^{-1}$ units. These fluxes include both the narrow and broad velocity components for the regions where two Gaussians were used.
\tablefoottext{a}{Intrinsic velocity dispersion after subtracting the instrumental width ($\sigma_{\rm inst}=39$\,km\,s$^{-1}$ at $\lambda_{\rm obs}$=4.7\micron; \citealt{Jakobsen2022}). For $^{12}$CO, the dispersions of the narrow ($\sigma_1$) and broad ($\sigma_2$) components are reported when two Gaussians were used for the fit.}
\tablefoottext{b}{The listed $^{13}$CO R(0) flux is likely contaminated by the $^{12}$CO P(11) transition.}}
\end{table*}

\subsection{ALMA rotational $^{12}$CO emission}\label{ss:12co_alma}

Using the ALMA observations of NGC~3256-S (Section~\ref{ss:data_alma}), we measured the fluxes of the CO(1--0)\,115.27\,GHz and \hbox{CO(2--1)}\,230.54\,GHz transitions in the selected regions (see Section~\ref{ss:sp_regions}). We extracted the spectra of the each region from the ALMA data cubes and integrated the line profile. The fluxes are listed in Table~\ref{tbl_line_fluxes_rot}. We also calculated the $r_{21}$ ratio between the fluxes of the CO(2--1) and \hbox{CO(1--0)} transitions in K\,km\,s$^{-1}$ units.

\begin{table*}
\caption{Fluxes of the $^{12}$CO rotational transitions}
\label{tbl_line_fluxes_rot}
\centering
\begin{small}
\begin{tabular}{lccccccccccccc}
\hline \hline
\\
CO transition & Frequency & \multicolumn{8}{c}{Region} \\[0.5ex]
\cline{3-10} \\[-2ex]
& GHz & N1 & N2 & N3 & N4 & N5 & S1 & S2 & S3 \\[0.5ex]
\hline
1--0 & 115.271 &  4.59  & 3.15 & 1.61 & 2.21 & 2.44 & 1.03 & 1.42 & 1.56 \\
2--1 & 230.538 &  16.72 & 9.55 & 5.00 & 6.07 & 7.81 & 4.69 & 5.88 & 5.68 \\
\hline
$r_{21}$\tablefootmark{a} & \nodata & 0.91 & 0.76 & 0.77 & 0.69 &  0.80 & 1.14& 1.04 & 0.91 \\
\hline
\end{tabular}
\end{small}
\tablefoot{The fluxes are in Jy\,km\,s$^{-1}$ units. The flux uncertainties are dominated by the calibration uncertainties which are 10\% for CO(1--0) and 20\% for CO(2--1).
\tablefoottext{a}{$r_{21}$ is the ratio between the CO(2--1) and CO(1--0) fluxes in K\,km\,s$^{-1}$.}}
\end{table*}

\section{Rovibrational CO $v$=1--0 emission model}\label{s:co_rovib_model}

As mentioned in Section~\ref{ss:spectral_co_maps}, in the outflow region of NGC~3256\,S, the CO $v$=1--0 emission can be excited by radiative pumping. The AGN is deeply dust embedded \citep{Ohyama2015} and this dust absorbs and re-emits part of the AGN radiation in the mid-IR at $\sim$4.7\micron. This re-emitted light behaves as a lamp that directly illuminates molecular clouds above and below the disk and excites the $v$=1 levels of CO through the absorption of $\sim$4.7\micron\ continuum photons. Then, the excited molecules decay to the fundamental $v$=0 levels emitting in the $v$=1--0 P- and R-branches (see sketch in Figure~\ref{fig_sketch}). For the P-branch transitions $J_{\rm up}-J_{\rm lo}=-1$, while for the R-branch transitions $J_{\rm up}-J_{\rm lo}=+1$, where $J$ is the rotational quantum number.

Under this geometry every absorbed $\sim$4.7\micron\ continuum photon results in the emission of a single P- or R-branch photon (according to the transition probabilities). In this case, the radiation field from the AGN is not intense enough to affect the excitation of the molecular cloud (see Section~\ref{ss:rp_heating}). Instead, the rovibrational CO emission acts as a probe of the physical conditions (temperature and column density) of the molecular gas in the illuminated clouds.

Here, we discuss ratios between P- and R-branch lines which trace the physical conditions of the molecular gas when the CO band is detected in emission.
We used in our analysis the CO molecular data from \citet{Goorvitch1994} and the collisional coefficients from \citet{Yang2010}.

\subsection{The P($J$+2)/R($J$) ratio}\label{ss:model_p2r}

\begin{figure*}
\centering
\includegraphics[width=17cm]{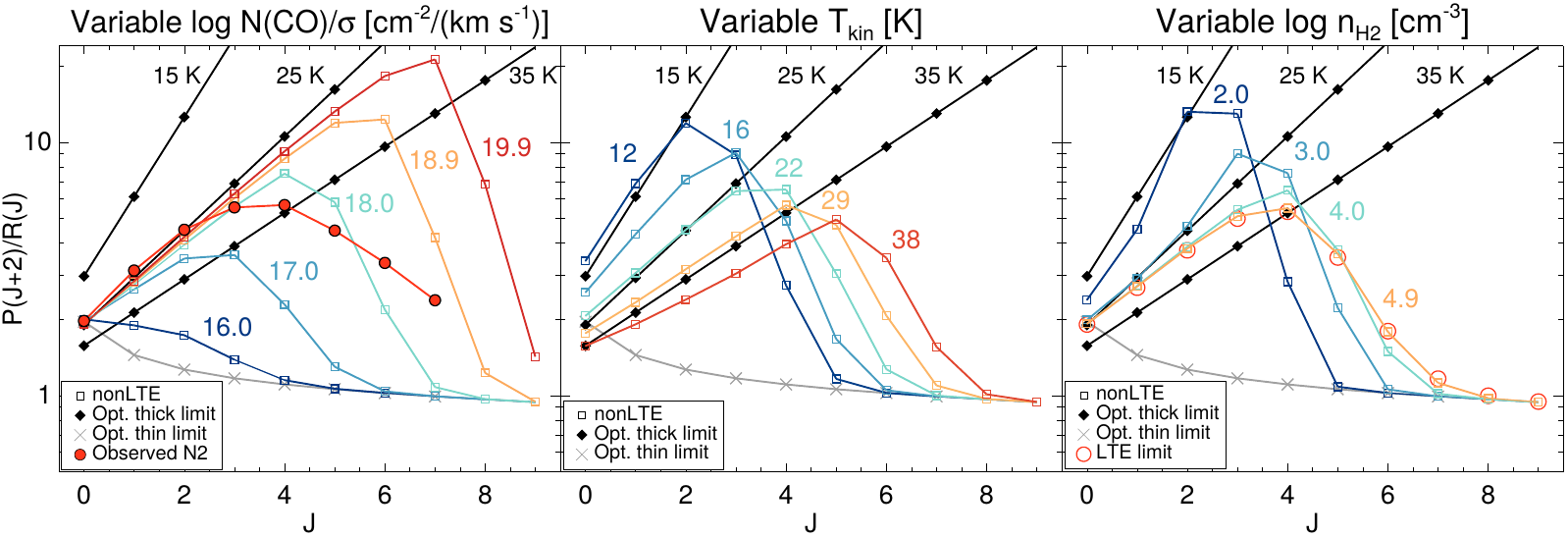}
\caption{Predicted P($J$+2)\slash R($J$)$\equiv$$f_J$ ratios for the $^{12}$CO $v$=1--0 band. The optically thin limit is  represented by the gray crosses (Section~\ref{ss:thin_limit}). The black diamonds show the optically thick limit for three rotational temperatures (15, 25, and 35\,K; Section~\ref{ss:thick_limit}). The dependence on $N({\rm CO})\slash \sigma_{\rm v}$ (left panel), $T_{\rm kin}$ (middle panel), and $n_{\rm H_2}$ (right panel) of the single component non-LTE models is indicated by the colored (blue to orange\slash red) empty squares. Only one parameter is varied in each panel (the colored numbers indicate the value of the varied parameter). The remaining two parameters are constant an equal to the following reference values: $N({\rm CO})\slash \sigma_{\rm v}$=10$^{17.7}$\,cm$^{-2}$\slash (km\,s$^{-1}$), $T_{\rm kin}$=25\,K, and $n_{\rm H_2}$=10$^4$\,cm$^{-3}$ (see Section~\ref{ss:nonLTE}).
The red filled circles in the left panel are the observed ratios in region N2. The red empty circles in the right panel represent the LTE limit for the reference values of $T_{\rm kin}$ and $N({\rm CO})\slash \sigma_{\rm v}$. The points are connected by lines to guide the eye.
\label{fig_pj2_models}}
\end{figure*}

We define the ratio between the P($J$+2) and R($J$) transitions as $f_J\equiv F_{{\rm P}(J+2)}\slash F_{{\rm R}(J)}$, where $F$ is the flux of each line. In this section, we determine the value of this ratio in the optically thin and thick limits and then we use non-local thermodynamic equilibrium (non-LTE) models to determine its dependence on the CO column density ($N({\rm CO})$), kinetic temperature ($T_{\rm kin}$), and number density ($n_{\rm H_2}$).

\subsubsection{Optically thin limit}\label{ss:thin_limit}

In the optically thin limit, the flux of a transition is
\begin{equation}
F = \Omega \frac{h \nu_{ul}}{4 \pi} A_{ul} N_{\rm up}
\label{eq_flux_thin}
\end{equation}
where $\Omega$ is the solid angle, $h$ the Planck constant, $N_{\rm up}$ the column density of the upper level, and $\nu_{ul}$ and $A_{ul}$ are the rest frequency and Einstein $A$-coefficient of the transition, respectively.

The P($J$+2) and R($J$) transitions share the same vibrationally excited upper level ($v_{\rm up}$=1, $J_{\rm up}=J+1$). Thus, their ratio only depends on molecular constants:
\begin{equation}
f_J^{\rm thin} = \nu_{{\rm P}(J+2)}\slash \nu_{{\rm R}(J)} \times A_{{\rm P}(J+2)}\slash A_{{\rm R}(J)}.
\end{equation}
The gray crosses in Figure~\ref{fig_pj2_models} show the optically thin limit ratios. The observed ratios, filled red circles in the left panel of Figure~\ref{fig_pj2_models}, clearly deviate from this limit.

\subsubsection{Optically thick limit}\label{ss:thick_limit}

Alternatively, if the CO rovibrational emission is optically thick, the flux of an emission line can be approximated by
\begin{equation}
F = \Omega \Delta {\rm v} \frac{2 h \nu_{ul}^4}{c^3} \frac{N_{\rm up}\slash g_{\rm up}}{N_{\rm lo}\slash g_{\rm lo}} 
\label{eq_flux_thick}
\end{equation}
where $c$ is the speed of light, $g$ is the level degeneracy, $2J+1$, of the upper and lower levels, $N_{\rm lo}$ the column density of the lower level, and $\Delta {\rm v}$ the width of the line. 
Therefore, the $f_J$ ratio is connected to the level population of the ground ($v$=0) CO states through this relation \citep{Sahai1985, GonzalezAlfonso2002}:
\begin{equation}
f_J^{\rm thick} = \left( \nu_{{\rm P}(J+2)}\slash \nu_{{\rm R}(J)} \right)^4  \times \frac{N_{0,J}/g_J}{N_{0,J+2}/g_{J+2}}
\label{eq_thick}
\end{equation}
where, $N_{v,J}$ is the population of the level with vibrational quantum number $v$ and rotational quantum number $J$, and $g_J$ is the level degeneracy.
In these equations it is assumed that the population of the vibrationally excited levels ($v$=1) is much smaller than that of the ground ($v$=0) levels ($N_{0,J}/N_{1,J} \gg 1$), which is reasonable for cold molecular regions detached from the mid-IR continuum source that excites the CO rovibrational emission by radiative pumping.

Equation~\ref{eq_thick} contains the ratio between the $v$=0 populations of the $J$ and $J+2$ levels, so it is also possible to connect the $f_J^{\rm thick}$ ratio and the excitation temperature, $T_{\rm rot}(J,J+2)$, within the  $v$=0 state:

\begin{equation}
f_J^{\rm thick} = \left( \nu_{{\rm P}(J+2)}\slash \nu_{{\rm R}(J)} \right)^4 \times \exp \left( \left( E_{0,J+2} - E_{0,J}  \right)\slash T_{\rm rot}(J,J+2) \right)
\label{eq_trot}
\end{equation}

where $E_{v,J}$ is the energy of the level with quantum numbers $v$ and $J$. Figure~\ref{fig_pj2_models} shows the optically thick limit for three temperatures (15--35\,K; black lines) which, in a semi-log plot, is a straight line whose slope and $y$-intercept depend on $T_{\rm rot}$.

For $J$<4, the observed $f_J$ (red circles in the left panel of Figure~\ref{fig_pj2_models}) approximately follow a straight line, which indicates that these transitions are optically thick.
For higher $J$, the observed ratios deviate from the optically thick limit toward the optically thin limit. This is expected since $N_{0,J}$, and the optical depth of the transitions, are expected to decrease with increasing $J$ for a sufficiently high $J$ and a finite column density as in real molecular clouds.

\subsubsection{Non-LTE models}\label{ss:nonLTE}

We produced a grid of non-LTE radiative transfer models to determine the dependence of the $f_J$ ratios on $T_{\rm kin}$ and $N({\rm CO})\slash \sigma_{\rm v}$, where $\sigma_{\rm v}$ is the velocity dispersion, and the H$_2$ density, $n_{\rm H_2}$. First, we used \textsc{RADEX} \citep{vanderTak2007} to determine the CO $v$=0 level populations for $T_{\rm kin}$ between 12 and 50\,K, $N({\rm CO})\slash \sigma_{\rm v}$ between 10$^{16.0}$ and 10$^{20}$\,cm$^{-2}$\slash (km\,s$^{-1}$), and $n_{\rm H_2}$ between 10$^{2}$ and 10$^{6}$\,cm$^{-3}$ with 50 log steps for each parameter (125~000 models in total).

Then, the flux of each transition is derived from the following radiative transfer equations: 

\begin{equation}
\sigma = \nu_{ul} \times \sigma_{\rm v}\slash c \label{eq_rt_1}
\end{equation}
\begin{equation}
\phi(\nu) = \frac{1}{(2\pi)^{1/2}\sigma}\times \exp \left(-\frac{\left(\nu - \nu_{ul} \right)^2}{2\sigma^2}\right)
\end{equation}
\begin{equation}
r = \frac{N_{\rm lo}\slash g_{\rm lo}}{N_{\rm up}\slash g_{\rm up}}
\end{equation}
\begin{equation}
\tau(\nu) = \frac{c^2 A_{ul}}{8\pi\nu^2} \times N_{\rm up} \times (r - 1) \times \phi(\nu)
\end{equation}
\begin{equation}
I(\nu) = \frac{2h \nu^3}{c^2 \left( r - 1 \right)} \times \left(1 - e^{-\tau(\nu)}\right) + B_\nu(T_{\rm CMB})\times \left( e^{-\tau(\nu)} - 1 \right) \label{eq_rt_5}
\end{equation}

where $N$ and $g$ the column density and degeneracy of the lower and upper levels of the transition, respectively, and $B_\nu(T_{\rm CMB})$ is the black body emission of the cosmic microwave background (CMB) at 2.73\,K. The CMB emission does not affect the rovibrational band, but it is needed to calculate the emission of the CO rotational transitions in Section~\ref{s:conv_factor}.
We note that for the rovibrational transitions $r-1\simeq r$ since $N_{\rm lo}\gg N_{\rm up}$. 
Equation~\ref{eq_rt_5} is the continuum subtracted emission, which we integrated over frequencies
to determine the intensities of the transitions.

By using these equations we ignore line overlaps. The separation between the observed rovibrational transitions is $>470$\,km\,s$^{-1}$ for $J$<10, so even for the broad outflow profiles (see Figure~\ref{fig_fit_model}), the effects of line overlapping are limited.
In Equation~\ref{eq_rt_5}, we also ignore the term accounting for the mid-IR background continuum in our line of sight since it has a low flux compared to that of the CO rovibrational lines.

For the $v$=0 level populations we used those calculated by \textsc{RADEX} while for the $v$=1 populations, we assumed that $N_{1,J}= k \times N_{0,J}$, where $k \ll 1$ is a constant scaling factor. 
We note that $k$ cancels out in the $f_j$ ratio, so its actual value is not relevant. Radiative excitation slightly alters the excitation temperature of the $v$=1 and $v$=0 levels, so the assumption $N_{1,J}= k \times N_{0,J}$ is not completely valid. However, when the excitation of the $v$=0 levels is dominated by collisions (see Section~\ref{ss:rp_heating}), a grid of full radiative transfer models, created using an updated version of the \citet{GonzalezAlfonso1998} code, indicates that this is an adequate approximation since deviations are small for optically thick lines ($<$30\,\%). For optically thin lines the deviations are also small resulting in rotational temperatures of the $v$=1 levels 3--5\,K higher than the rotational temperature of the $v$=0 levels.

Figure~\ref{fig_pj2_models} shows the $f_{\rm J}$ ratios predicted by these models.
The first panel shows the dependence on $N({\rm CO})\slash \sigma_{\rm v}$ for fixed $T_{\rm kin}$ and $n_{\rm H_2}$. The ratios remain close to the optically thick limit up to a given $J$ which increases with increasing $N({\rm CO})\slash \sigma_{\rm v}$. This is because the optical depth of the transitions, $\tau$, increases with $N({\rm CO})\slash \sigma_{\rm v}$.

The middle panel of Figure~\ref{fig_pj2_models} shows the dependence on $T_{\rm kin}$. For fixed $N({\rm CO})\slash \sigma_{\rm v}$ and $n_{\rm H_2}$, the low-$J$ ratios (e.g., $f_{\rm 0}$) increase with decreasing $T_{\rm kin}$. Thus, the $J$ at which the observed ratios deviate from the optically thick limit together with the low-$J$ ratios can constrain both the CO column density and $T_{\rm kin}$.

The dependence on $n_{\rm H_2}$ is presented in the right panel of Figure~\ref{fig_pj2_models}. For $n_{\rm H_2}\leq10^3$\,cm$^{-3}$, the $f_{\rm J}$ ratios have a positive curvature before reaching the maximum value. 
For $n_{\rm H_2}>10^4$\,cm$^{-3}$, the ratios are almost equal to the LTE limit and are not sensitive to $n_{\rm H_2}$.
This positive curvature is only observed in one of the regions studied here (N5) where it can help constraining $n_{\rm H_2}$. However, in most cases this curvature is not present and we can only estimate that their $n_{\rm H_2}$ is larger than $10^{3-4}$\,cm$^{-3}$.

In Section~\ref{s:conv_factor}, we use these non-LTE models to estimate the CO-to-H$_2$ conversion factor.

\subsection{Collisions vs. IR radiative pumping}\label{ss:rp_heating}

In this section, we investigate if the excitation of the pure rotational $v$=0 levels of CO is affected by the IR radiative pumping in these regions as occurs for other molecules when the ratio between the intensity of the radiation and $n_{\rm H_2}$ is high enough (e.g., HNC and HCN; \citealt{Aalto2007, Imanishi2017}).

The de-excitation of the $v$=1 CO levels is dominated by the spontaneous emission of photons in the $v$=1--0 band, which has transitions probabilities much higher than the rotational transitions within two $v$=1 levels. Moreover, the densities expected in molecular outflows are well below the critical density  ($n^{\rm cr}_{\rm H_2}$$>$10$^{17}$\,cm$^{-3}$ at 160\,K for collisions with H$_2$; see \citealt{GonzalezAlfonso2002}) of these transitions. Therefore, collisional and rotational de-excitation of the $v$=1 levels can be neglected.

In this context, it is reasonable to assume that every absorbed $\sim$4.7\micron\ continuum photon is re-emitted as a single photon in the $v$=1--0 P- or R-branch. Thus, assuming that the CO $v$=1--0 emission is isotropic, we transformed the observed CO \hbox{$v$=1--0} fluxes (Table~\ref{tbl_line_fluxes}) into the rate of \hbox{$v$=1--0} emitted photons, that is, the IR radiative pumping rate $R_{\rm IR}$ (e.g., \citealt{GonzalezAlfonso2022}). We find the rates in the range (0.3--4.1)$\times$10$^{51}$\,s$^{-1}$.

To estimate the rate of collisions, first we obtained the number of CO molecules, $\mathcal{N}_{\rm CO}$, in each region from the \hbox{CO(1--0)}\,115.27\,GHz luminosity (Table~\ref{tbl_line_fluxes_rot}) assuming an $\alpha_{\rm CO}$ factor and a CO abundance relative to H$_2$, $[$CO\slash H$_2]$. 
The rate of collisions is $\gamma_{ul} n_{\rm H_2} \mathcal{N}_{\rm CO}$ where $\gamma_{ul}$ are the collisional de-excitation coefficients with H$_2$ for the lowest CO $v$=0 rotational levels ((0.35--1.6) $\times$10$^{-10}$\,cm$^3$\,s$^{-1}$; \citealt{Yang2010}).

We find that the ratio between the IR radiative pumping rate and the rate of collisions is
\begin{multline}
\frac{R_{\rm IR}}{\gamma_{ul} n_{\rm H_2} \mathcal{N}_{\rm CO}} = (0.2-5.8)\times10^{-3}  \\
\times \frac{0.8 M_\odot\,{\rm (K\,km\,s^{-1}\,pc^2})^{-1}}{\alpha_{\rm CO}} \times 
\frac{10^4\,{\rm cm^{-3}}}{n_{\rm H_2}} \times \frac{3.2 \times 10^{-4}}{{\rm [CO\slash H_2]}}
\end{multline}

Thus, for the ULIRG-like $\alpha_{\rm CO}$ (0.8\,$M_\odot$\,(K\,km\,s$^{-1}$\,pc$^2$)$^{-1}$; \citealt{Bolatto2013}) usually assumed for outflows, the excitation of the rotational CO levels is dominated by collisions for densities $n_{\rm H_2}>$10$^2$\,cm$^{-2}$. 
Even if the $\alpha_{\rm CO}$ of outflows is down to a factor of two lower than the ULIRG-like factor (see Section~\ref{s:conv_factor}), the contribution from radiative pumping to the rotational excitation of CO would be two times higher but still not dominant ($<$2\%).

\begin{figure*}
\centering
\includegraphics[width=0.28\textwidth]{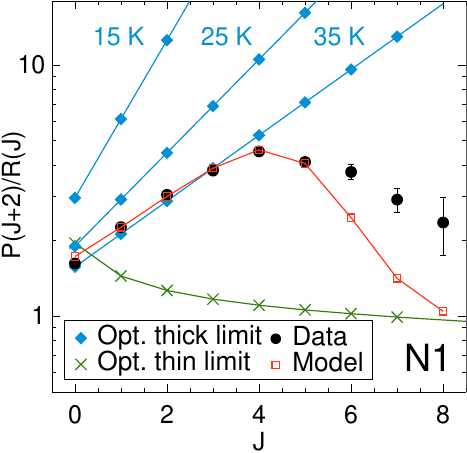}
\includegraphics[width=0.28\textwidth]{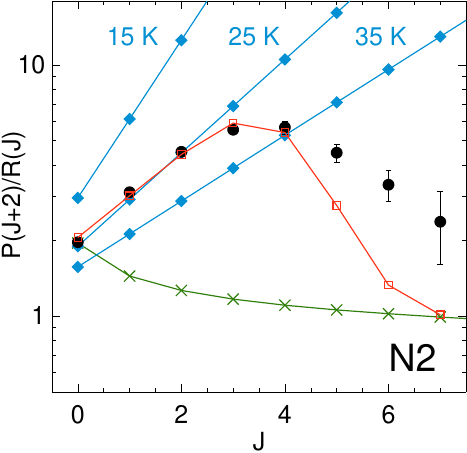}
\includegraphics[width=0.28\textwidth]{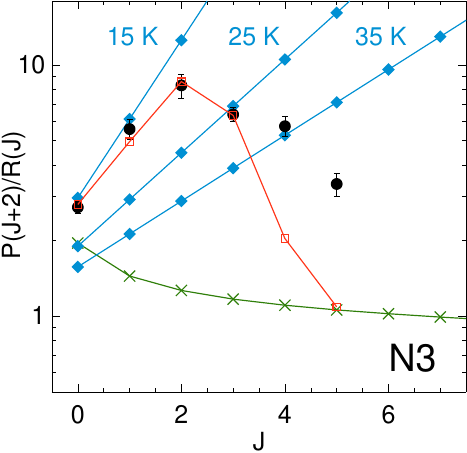}
\includegraphics[width=0.28\textwidth]{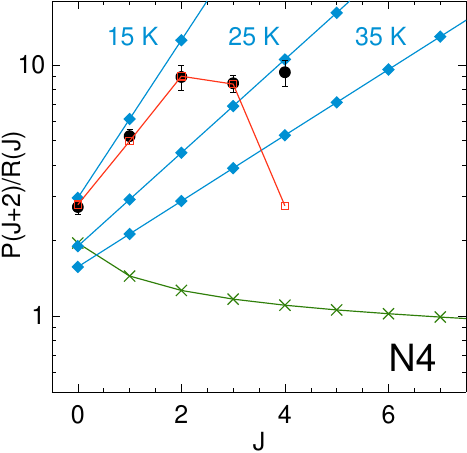}
\includegraphics[width=0.28\textwidth]{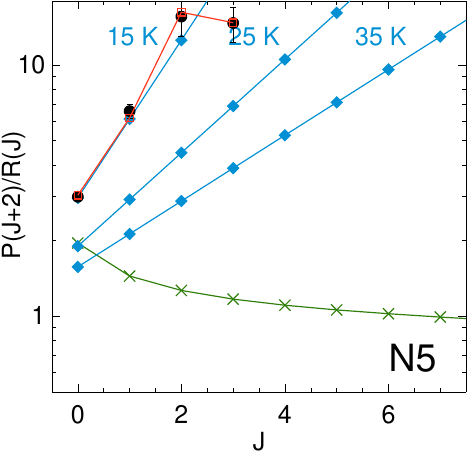}
\includegraphics[width=0.28\textwidth]{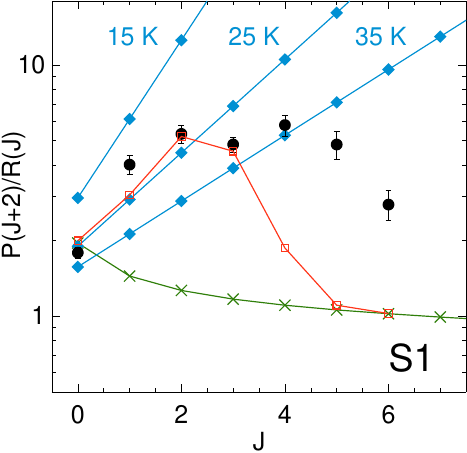}
\includegraphics[width=0.28\textwidth]{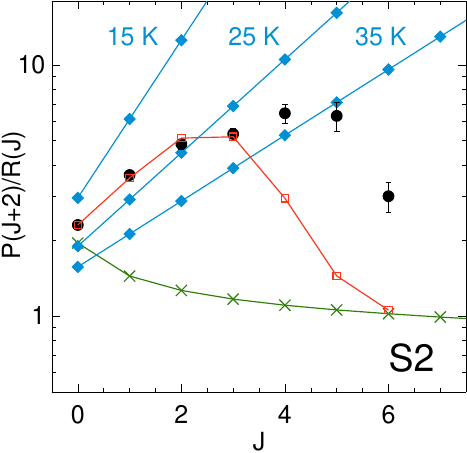}
\includegraphics[width=0.28\textwidth]{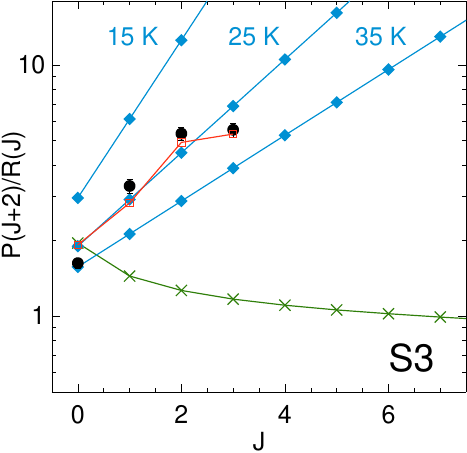}
\caption{Ratio between the P($J$+2) and R($J$) transitions of the CO $v$=1--0 band for the selected regions (black circles). The best-fit single component model is represented by the red empty squares (see Section~\ref{ss:one_comp}). The parameters of the best-fit models are listed in Table~\ref{tbl_single_component}.
The blue diamonds and green crosses show the optically thick and thin limits, respectively (see Sections~\ref{ss:thin_limit} and \ref{ss:thick_limit}). The points are connected by lines to guide the eye.
\label{fig_one_comp}}
\end{figure*}

\section{CO-to-H$_2$ conversion factor}\label{s:conv_factor}

It is possible to determine the CO-to-H$_2$ conversion factor ($\alpha_{\rm CO}=M_{\rm mol}\slash L_{\rm CO(1-0)}$) if the physical conditions, $T_{\rm kin}$ and $N({\rm CO})\slash \sigma_{\rm v}$, of the gas are known. 
Using Equations~\ref{eq_rt_1}-\ref{eq_rt_5}, we derived the integrated intensity, $W$(CO(1--0)), of the CO(1--0)\,115.27\,GHz transition in K\,km\,s$^{-1}$. $N({\rm H_2})$ is calculated as $(N({\rm CO})\slash \sigma_{\rm v})\times \sigma_{\rm v} \slash[$CO\slash H$_2]$. Then, the ratio $N({\rm H_2})$\slash $W$\hbox{(CO(1--0))}, known as $X_{\rm CO}$, is converted to $\alpha_{\rm CO}$ multiplying the H$_2$ mass by 1.36 to account for helium and other elements \citep{Bolatto2013}.
Knowing the exact value of $\sigma_{\rm v}$ is not critical since, in the optically thick limit,  $W$(CO(1--0))\,$\propto T_{\rm b}\times \sigma_{\rm v}$, where $T_{\rm b}$ is the peak brightness temperature, so it cancels out in the $X_{\rm CO}$ definition.

We discuss three different methods to measure $\alpha_{\rm CO}$ using the CO $v$=1--0 band. We used a single component non-LTE model (see Section~\ref{ss:nonLTE}) to fit the \hbox{P($J$+2)/R($J$)} ratios (Section~\ref{ss:one_comp}). This single component models fail to reproduce the higher-$J$ ratios, so we included a second component in Section~\ref{ss:two_comp}. Finally, for the regions where the $^{13}$CO $v$=1--0 band is well detected, we used it to constrain the excitation temperature of the $^{12}$CO $v$=0 levels (Section~\ref{ss:13co_model}).

\subsection{Single component model}\label{ss:one_comp}

Figure~\ref{fig_one_comp} shows the best-fit for each region using a single non-LTE model. A single component is not able to simultaneously reproduce the low- and high-$J$ \hbox{P($J$+2)/R($J$)} ratios.
We focus on fitting the low-$J$ transitions since these are likely tracing the bulk of the molecular gas mass.

The higher-$J$ ratio used in the fit for each region is set as the maximum $J$ that can be reproduced by a single component model (i.e., when $\chi^2$ does not significantly increases when adding that ratio). This single component model can fit between four and six ratios. The physical conditions of the gas derived for each region are listed in Table~\ref{tbl_single_component}.

\begin{table*}
\caption{Single component models}
\label{tbl_single_component}
\centering
\begin{small}
\begin{tabular}{lcccccccccc}
\hline \hline
\\
Region & $T_{\rm kin}$ & $\log N({\rm CO})\slash \sigma_{\rm v}$ & $\log n_{\rm H_2}$ & $\alpha_{\rm CO}\slash\frac{3.2 \times10^{-4}}{[{\rm CO}\slash {\rm H_2}]}$ & $r_{21}$\tablefootmark{b} & $r_{31}$\tablefootmark{b} \\[1ex]
 & K & cm$^{-2}$ (km\,s$^{-1}$)$^{-1}$ & cm$^{-3}$ & $M_\odot$ (K km\,s$^{-1}$\,pc$^2$)$^{-1}$ \\
\hline
N1 & 31.0 $\pm$ 1.3 & 17.71 $\pm$ 0.12 & $>$3.9 & 0.34 $\pm$ 0.07 & 1.23 $\pm$ 0.03 & 1.24 $\pm$ 0.05 \\
N2 & 22.8 $\pm$ 0.9 & 17.63 $\pm$ 0.12 & $>$3.4 & 0.40 $\pm$ 0.12 & 1.11 $\pm$ 0.03 & 1.05 $\pm$ 0.05 \\
N3 & 14.9 $\pm$ 1.1 & 17.47 $\pm$ 0.08 & $>$2.9 & 0.40 $\pm$ 0.05 & 1.05 $\pm$ 0.03 & 0.85 $\pm$ 0.05 \\
N4 & 14.9 $\pm$ 0.4 & 17.63 $\pm$ 0.08 & $>$3.0 & 0.53 $\pm$ 0.08 & 1.00 $\pm$ 0.03 & 0.85 $\pm$ 0.05 \\
N5 & 14.1 $\pm$ 0.8 & 17.96 $\pm$ 0.25 & $>$2.5 & 1.09 $\pm$ 0.56 & 1.00 $\pm$ 0.03 & 0.75 $\pm$ 0.05 \\
S1 & 27.0 $\pm$ 5.0 & $>$17.0          & $>$2.5 & $>$0.2          & 1.05 $\pm$ 0.12 & 1.05 $\pm$ 0.10 \\
S2 & 18.7 $\pm$ 0.8 & 17.31 $\pm$ 0.09 & $>$3.3 & 0.26 $\pm$ 0.02 & 1.17 $\pm$ 0.03 & 1.05 $\pm$ 0.05 \\
S3 & 26.2 $\pm$ 5.0 & $>$16.9          & $>$3.1 & $>$0.2          & 1.05 $\pm$ 0.12 & 0.95 $\pm$ 0.10 \\
Mean   & \nodata\tablefootmark{a} & 17.62 $\pm$ 0.24 & \nodata & 0.50 $\pm$ 0.30 & 1.08 $\pm$ 0.09 & 0.97 $\pm$ 0.16 \\
Median & \nodata\tablefootmark{a} & 17.63 $\pm$ 0.20 & \nodata & 0.40 $\pm$ 0.15 & 1.05 $\pm$ 0.07 & 1.00 $\pm$ 0.15 \\
\hline
\end{tabular}
\end{small}
\tablefoot{Parameters of the best-fit single component model for the selected regions (Section~\ref{ss:one_comp}).
\tablefoottext{a}{The mean and median $T_{\rm kin}$ of the regions is not indicated since there is a temperature gradient with decreasing $T$ for increasing distances from the nucleus (Figure~\ref{fig_temp_map}).}
\tablefoottext{b}{Ratio between the rotational transitions \hbox{CO(2--1)}\slash CO(1--0) ($r_{21}$) and CO(3--2)\slash CO(1--0) ($r_{31}$) in K\,km\,s$^{-1}$ units.
}
}
\end{table*}

\begin{figure}
\centering
\includegraphics[width=0.38\textwidth]{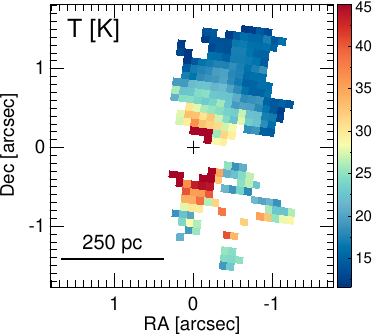}
\caption{Rotational temperature of the molecular gas derived from the ratio between the CO $v$=1--0 P(2) and R(0) transitions assuming the optically thick limit (Equation~\ref{eq_thick}). Only positions where both the R- and P- branches are in emission are considered. The black cross marks the position of the southern nucleus of NGC~3256.
\label{fig_temp_map}}
\end{figure}

$T_{\rm kin}$ varies between 14 and 35\,K. For these regions, we find a temperature gradient with decreasing $T_{\rm kin}$ at further distances from the nucleus. To investigate this gradient, we created a 2D map of rotational temperature using the $f_0$ ratio measured in each spaxel (Section~\ref{ss:spectral_maps}) assuming the optically thick limit (Equation~\ref{eq_thick}). In this limit, the rotational temperature is equal to $T_{\rm kin}$ if the low-$J$ levels are thermalized.
Figure~\ref{fig_temp_map} clearly shows, specially at the northern outflow region, that the molecular gas cools down from $>$40\,K at the central 100\,pc to $<$15\,K at $\sim$250\,pc. This result is also consistent with the low kinetic temperature ($\sim$9\,K) in the molecular outflow compared to the nuclear gas temperature ($\sim$40\,K) measured in the LIRG ESO~320-G030 at similar spatial scales \citep{Pereira2020}.

The number density $n_{\rm H_2}$ is not well constrained in general with lower limits around 10$^{2.5-4}$\,cm$^{-3}$. $N({\rm CO})\slash \sigma_{\rm v}$ is better constrained and it varies between $>$10$^{16.9}$ and 10$^{18.0}$\,cm$^{-2}$ (km\,s$^{-1}$)$^{-1}$.

From these physical conditions, we also calculated the $\alpha_{\rm CO}$ conversion factor as described above in Section~\ref{s:conv_factor}. We find $\alpha_{\rm CO}$ between (0.26--0.53)$\times\frac{3.2 \times10^{-4}}{[{\rm CO}\slash {\rm H_2}]}$\,$M_\odot$ (K km\,s$^{-1}$\,pc$^2$)$^{-1}$ with a median (mean) of 0.40 $\pm$ 0.15 (0.50 $\pm$ 0.30). This is $\sim$10 times lower than the Galactic $\alpha_{\rm CO}$ (4.3\,$M_\odot$ (K km\,s$^{-1}$\,pc$^2$)$^{-1}$; e.g., \citealt{Bolatto2013}), and also $\sim$2 times lower than the ULIRG $\alpha_{\rm CO}$ (0.8\,$M_\odot$ (K km\,s$^{-1}$\,pc$^2$)$^{-1}$) which is typically used to estimate the mass of cold molecular outflows (e.g., \citealt{Cicone2014, Fiore2017, Pereira2018, Lutz2020, Lamperti2022}).

It is also common to observe alternative low-$J$ CO rotational transitions, like the 2--1 or the 3--2 at 230.54 and 345.80\,GHz, respectively, instead of the CO(1--0). In these cases, the observed flux must be converted to an equivalent CO(1--0) flux before using the $\alpha_{\rm CO}$ factor by assuming a CO(2--1) to CO(1--0) ratio ($r_{21}$) or CO(3--2) to CO(1--0) ratio ($r_{32}$). From the models, we also determined the expected $r_{21}$ and $r_{31}$ ratios. The mean ratios are 1.08 $\pm$ 0.09 and 0.97 $\pm$ 0.16, respectively. The measured $r_{21}$ in the outflow regions are between 0.69 and 1.14 (Table~\ref{tbl_line_fluxes_rot}), with a mean ratio of 0.88 $\pm$ 0.15, that is in good agreement with the models.

\subsection{Two component model}\label{ss:two_comp}

\begin{figure*}
\centering
\includegraphics[width=0.28\textwidth]{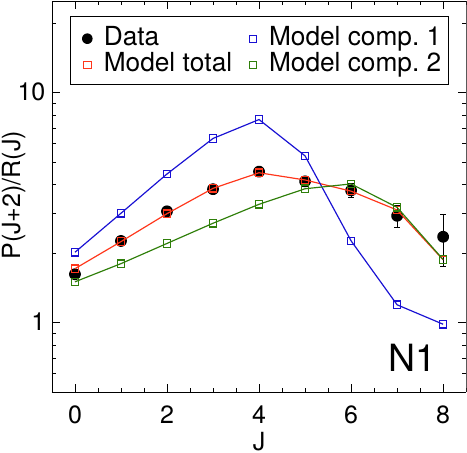}
\includegraphics[width=0.28\textwidth]{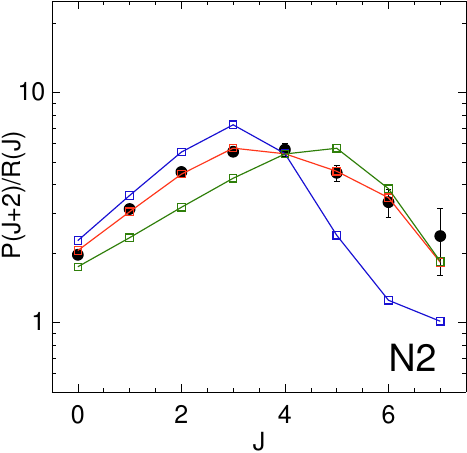}
\includegraphics[width=0.28\textwidth]{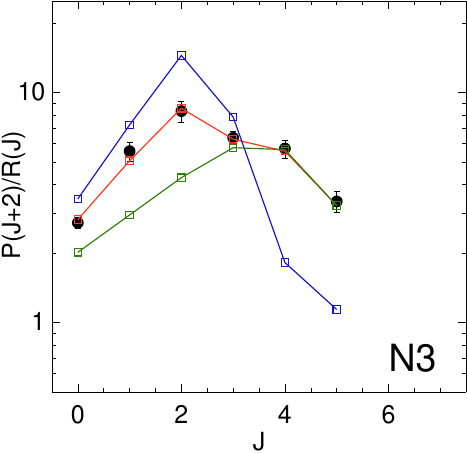}
\includegraphics[width=0.28\textwidth]{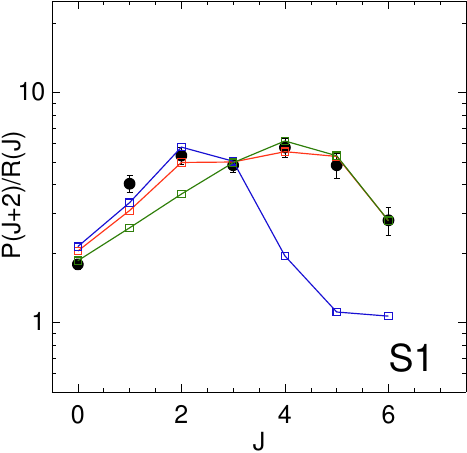}
\includegraphics[width=0.28\textwidth]{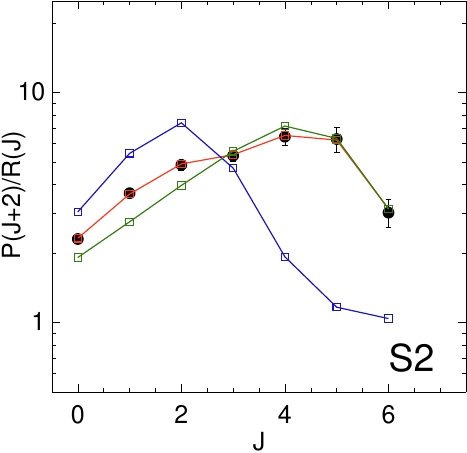}
\caption{Ratio between the P($J$+2) and R($J$) transitions of the CO $v$=1--0 band for the selected regions ( filled black circles). The best-fit two component model is shown as empty red squares (see Section~\ref{ss:two_comp}). The blue and green empty squares represent components 1 (``cold'' ) and 2 (``warmer'') of the best-fit model, respectively. The parameters of the best-fit model are listed in Table~\ref{tbl_two_component}. The points are connected by lines to guide the eye.
\label{fig_two_comp}}
\end{figure*}

\begin{table*}
\caption{Two-component models}
\label{tbl_two_component}
\centering
\begin{small}
\begin{tabular}{lcccccccccccccc}
\hline \hline
\\
Region & \multicolumn{2}{c}{Component 1} & & \multicolumn{2}{c}{Component 2} & $\alpha_{\rm CO}\slash\frac{3.2 \times10^{-4}}{[{\rm CO}\slash {\rm H_2}]}$ & $r_{21}$\tablefootmark{a} & $r_{31}$\tablefootmark{a} \\
\cline{2-3}  \cline{5-6} \\[-2ex]
 & $T_{\rm kin}$ & $\log N({\rm CO})\slash \sigma_{\rm v}$ & & $T_{\rm kin}$ & $\log N({\rm CO})\slash \sigma_{\rm v}$  & \\[1ex]
 & K & cm$^{-2}$ (km\,s$^{-1}$)$^{-1}$  & & K & cm$^{-2}$ (km\,s$^{-1}$)$^{-1}$  & $M_\odot$ (K km\,s$^{-1}$\,pc$^2$)$^{-1}$ \\
\hline
N1 & 20 $\pm$ 7 & 18.5 $\pm$ 0.9 & & 39 $\pm$ 7 & 17.9 $\pm$ 0.3 & 0.61$^{+1.06}_{-0.21}$ & 1.23 $\pm$ 0.09 & 1.21 $\pm$ 0.09 \\
N2 & 17 $\pm$ 3 & 18.2 $\pm$ 0.7 & & 32 $\pm$ 10 & 17.8 $\pm$ 0.5 & 0.71$^{+1.23}_{-0.25}$ & 1.13 $\pm$ 0.10 & 1.03 $\pm$ 0.13 \\
N3 & 14 $\pm$ 1 & 17.6 $\pm$ 0.6 & & 23 $\pm$ 12 & 17.8 $\pm$ 0.8 & 0.53$^{+0.73}_{-0.19}$ & 1.04 $\pm$ 0.06 & 0.88 $\pm$ 0.10 \\
S1 & 19 $\pm$ 5 & 18.1 $\pm$ 0.9 & & 31 $\pm$ 9 & 17.7 $\pm$ 0.7 & 0.61$^{+1.32}_{-0.21}$ & 1.13 $\pm$ 0.12 & 1.03 $\pm$ 0.13 \\
S2 & 15 $\pm$ 2 & 17.7 $\pm$ 0.9 & & 25 $\pm$ 8 & 17.9 $\pm$ 0.6 & 0.61$^{+1.06}_{-0.21}$ & 1.10 $\pm$ 0.08 & 0.95 $\pm$ 0.12 \\

Mean & 17 $\pm$ 5 & 18.0 $\pm$ 0.8 & & 30 $\pm$ 10 & 17.8 $\pm$ 0.4 & 0.62 $\pm$ 0.20 & 1.13 $\pm$ 0.07 & 1.02 $\pm$ 0.12 \\
Median & 17 $\pm$ 4 & 18.1 $\pm$ 0.6 & & 31 $\pm$ 8 & 17.8 $\pm$ 0.3 & 0.61 $\pm$ 0.17 & 1.13 $\pm$ 0.04 & 1.03 $\pm$ 0.12 \\
\hline
\end{tabular}
\end{small}
\tablefoot{Parameters of the best-fit two-component model for the selected regions (Section~\ref{ss:two_comp}). $\alpha_{\rm CO}$ is weighted by the intensity of the CO(1--0)\,115.27\,GHz intensity of each component.
\tablefoottext{a}{Ratio between the rotational transitions \hbox{CO(2--1)}\slash CO(1--0) ($r_{21}$) and CO(3--2)\slash CO(1--0) ($r_{31}$) in K\,km\,s$^{-1}$ units.}
}
\end{table*}

The single component models are not able to reproduce all the $f_J$ ratios simultaneously. In this section, we add a second component to the models of regions with six or more $f_J$ ratios detected where a single component clearly fails to reproduce the observations. To find the best-fit model, we calculated the $\chi^2$ for all the possible combinations of two models of the grid leaving only a scaling factor between the two models (ratio between their solid angles) as a free parameter. To limit the number of combinations, we only considered three number densities: 10$^3$, 10$^4$, and 10$^5$\,cm$^{-3}$ given that the CO $v$=1--0 band is not very sensitive to $n_{\rm H_2}$. In total, we explored 56\,million combinations of models for each region.

The best-fit models are shown in Figure~\ref{fig_two_comp}. With two components, it is now possible to fit all the observed ratios. 
To determine the likelihood function of the parameters, we assigned a probability to each model of $\exp(-\chi^2\slash 2)$ and then calculated the marginalized likelihoods for each parameter.
The parameter values are listed in Table~\ref{tbl_two_component}. We find that the best-fit models include a ``cold'' (Component 1), with a mean $T_{\rm kin}$ of 17\,K, and a ``warmer'' component (Component 2), with a mean $T_{\rm kin}$ of 30\,K. The $N({\rm CO})\slash \sigma_{\rm v}$ of the cold component is $\sim$10$^{18.0}$\,cm$^{-2}$ (km\,s$^{-1}$)$^{-1}$, which is $\sim$0.2\,dex higher than that of the warm component.
We calculated the $\alpha_{\rm CO}$ weighted by the intensity of the CO(1--0)\,115.27\,GHz intensity of each component. The $\alpha_{\rm CO}$ are between (0.5--0.7)$\times\frac{3.2 \times10^{-4}}{[{\rm CO}\slash {\rm H_2}]}$\,$M_\odot$ (K km\,s$^{-1}$\,pc$^2$)$^{-1}$ with a median value of 0.61 $\pm$ 0.17. These $\alpha_{\rm CO}$ are 1.3--2.3 times higher than the $\alpha_{\rm CO}$ derived using a single component model for the individual regions. It is $>$7 times lower than the Galactic $\alpha_{\rm CO}$, and slightly lower ($\sim$30\%) than the ULIRG-like $\alpha_{\rm CO}$.
We find that the $r_{21}$ and $r_{31}$ ratios are consistent, within the uncertainties, with those derived using a single component model (Section~\ref{ss:one_comp}).

\subsection{Optically thin $^{13}$CO $v$=1--0 band}\label{ss:13co_model}

The P-R line ratio $f_0$ indicates that the $^{13}$CO \hbox{$v$=1--0} band is optically thin (Section~\ref{ss:13cothin}). Therefore, it can be used to estimate the mass of the molecular clouds in a similar way to the standard method for the sub-mm\slash mm $^{13}$CO rotational transitions (e.g., \citealt{Papadopoulos2012b}).

From Equation~\ref{eq_flux_thin}, for $^{13}$CO in the optically thin limit, the population of the upper level is:
\begin{equation}
\Omega N^{\rm ^{13}CO}_{\rm up} = \frac{4 \pi}{h \nu_{ul} A_{ul}}  F^{\rm ^{13}CO} .
\label{eq_13co_pop}
\end{equation}

For $^{12}$CO in the optically thick limit using Equation~\ref{eq_flux_thick} we find:

\begin{equation}
\frac{\Omega N^{\rm ^{12}CO}_{\rm up}}{N^{\rm ^{12}CO}_{\rm lo} \slash \Delta {\rm v}} =  \frac{c^3}{2 h \nu_{ul}^4} \frac{g_{\rm up}}{g_{\rm lo}}  F^{\rm ^{12}CO} .
\label{eq_12co_pop}
\end{equation}

Thus, assuming that $\Omega$ and the excitation temperature of the $v$=1 and $v$=0 levels is the same for $^{12}$CO and $^{13}$CO, the ratio between Equations~\ref{eq_13co_pop} and \ref{eq_12co_pop} gives for the  left-hand side of the ratio (${N^{\rm ^{12}CO}_{\rm lo} \slash \Delta {\rm v}}$)$\slash [^{12}$CO\slash $^{13}$CO$]$. From the $^{13}$CO and $^{12}$CO R(0) transitions, we estimate $N^{\rm ^{12}CO}_{{\rm v}=0,J=0}$ \slash $\Delta {\rm v}$. To determine the intensity of the CO(1--0)\,115.27\,GHz transition using Equations~\ref{eq_rt_1}--\ref{eq_rt_5}, we need the population of the $N^{\rm ^{12}CO}_{{\rm v}=0,J=1}$ level, which we estimated based on the excitation temperature of the low-$J$ levels, 15--30\,K, of these regions  derived in Section~\ref{ss:one_comp} (see also Table~\ref{tbl_single_component} and Figure~\ref{fig_temp_map}). The population of the remaining $N^{\rm ^{12}CO}_{{\rm v}=0,J}$ levels is also estimated using the Section~\ref{ss:one_comp} excitation temperatures.
We also need to transform the observed $\Delta {\rm v}$ (FWHM) into the $\sigma_{\rm v}$ used in Equation~\ref{eq_rt_1}. To do this, we first divided $\Delta {\rm v}$ by $\sim$2.355 to account for the ratio between FWHM and $\sigma$. The observed $\Delta {\rm v}$ also depends on the optical depths of the $^{12}$CO $v$=1--0 transitions which are unknown. To account for this uncertainty, we divided $\Delta {\rm v}$ by an extra factor between 1.2 and 2.6 (valid for $\tau$ between 1--100).

Then, estimating $M_{\rm mol}$ as described at the beginning of Section~\ref{s:conv_factor}, we derived the $\alpha_{\rm CO}$ conversion factor for these eight regions (Table~\ref{tbl_13co_models}). The $\alpha_{\rm CO}$ values are between $(0.15-0.59)\times\frac{3.2 \times10^{-4}}{[{\rm CO}\slash {\rm H_2}]}\times\frac{[{\rm ^{12}CO}\slash {\rm ^{13}CO}]}{60}$\,$M_\odot$ (K km\,s$^{-1}$\,pc$^2$)$^{-1}$ with a median range of 0.19--0.36. The median $\alpha_{\rm CO}$ are 1.1--2.1 times lower than those derived using the single component non-LTE model (Section~\ref{ss:one_comp}) and 1.7--3.2 times lower than those obtained from the two-component models (Section~\ref{ss:two_comp}). These $\alpha_{\rm CO}$ derived from $^{13}$CO are also more than 2 and 10 times lower than the ULIRG-like and Galactic $\alpha_{\rm CO}$ factors, respectively. We note that the $\alpha_{\rm CO}$ derived using the $^{13}$CO band
depend on the $^{12}$CO abundance and the $^{13}$CO to $^{12}$CO abundance ratio in the outflow, which are both uncertain. Therefore, these $\alpha_{\rm CO}$ are more uncertain than those derived using the single and two component methods (Sections~\ref{ss:one_comp} and \ref{ss:two_comp}) which only depend on the $^{12}$CO abundance.

\begin{table}
\caption{$^{13}$CO models}
\label{tbl_13co_models}
\centering
\begin{small}
\begin{tabular}{lcccccccccc}
\hline \hline
\\
Region & $T_{\rm kin}$ & $\alpha_{\rm CO}\slash (\frac{3.2 \times10^{-4}}{[{\rm CO}\slash {\rm H_2}]}\times\frac{[{\rm ^{12}CO}\slash {\rm ^{13}CO}]}{60})$  \\[1ex]
 & K  & $M_\odot$ (K km\,s$^{-1}$\,pc$^2$)$^{-1}$ \\
\hline
N1 & 30 & 0.15--0.32 \\
N2 & 22 & 0.19--0.44 \\
N3 & 15 & 0.23--0.56 \\
N4 & 15 & 0.19--0.40 \\
N5 & 14 & 0.18--0.32 \\
S1 & 27 & 0.20--0.31 \\
S2 & 19 & 0.17--0.27 \\
S3 & 26 & 0.20--0.59 \\
Mean & \nodata & 0.19--0.40\\
Median &\nodata & 0.19--0.36\\
\hline
\end{tabular}
\end{small}
\tablefoot{Parameters of the best-fit $^{13}$CO models. The uncertainty range for $\alpha_{\rm CO}$ is related to the range of correction factors needed to take into account the unknown optical depth of the $^{12}$CO $v$=1--0 transitions (see Section~\ref{ss:13co_model}).}
\end{table}

\section{Discussion}

We found that the $\alpha_{\rm CO}$ factors in the molecular outflow are 1.3--2 times lower than the ULIRG conversion factor and 6--8 times lower than the Galactic factor (assuming \hbox{$[$CO\slash H$_2$$]$}=3.2$\times$10$^{-4}$; Section~\ref{s:conv_factor}). We discuss here several options to explain the origin of this reduced $\alpha_{\rm CO}$ in outflows.

For individual molecular clouds, the existence of an $\alpha_{\rm CO}$ factor, which relates the total molecular mass and the luminosity of the CO(1--0)\,115.27\,GHz line, relies on several assumptions: (1) the CO(1--0) emission is optically thick; (2) the molecular clouds are virialized; and (3) the mass is dominated by H$_2$ (see e.g., \citealt{Bolatto2013}). Under these conditions 
\begin{equation}
\alpha_{\rm CO} \propto \frac{\sqrt{n_{\rm H_2}}}{T_{\rm kin}}. \label{eq_alphaco}
\end{equation}

All these assumptions seem to be valid for molecular clouds in the Milky Way and nearby spiral galaxies where  $\alpha_{\rm CO}$=3.1--4.3\,$M_\odot\,{\rm (K\,km\,s^{-1}\,pc^2})^{-1}$ (e.g., \citealt{Sandstrom2013, Bolatto2013}). The temperatures (15--40\,K) and densities ($>$10$^3$\,cm$^{-3}$) we measured in the molecular outflow do not significantly differ from those observed in the molecular clouds found in disks. Thus, the $n_{\rm H_2}$ and $T_{\rm kin}$ dependencies shown in Equation~\ref{eq_alphaco} would not explain the low $\alpha_{\rm CO}$ in the outflow.

\citet{Dasyra2016} found that the rotational CO emission in the jet-driven outflow of IC~5063 is optically thin, so $\alpha_{\rm CO}$ reaches a lower limit of 0.11\,$M_\odot\,{\rm (K\,km\,s^{-1}\,pc^2})^{-1}$ (for $T_{\rm ex}$=30\,K and \hbox{$[$CO\slash H$_2$$]$}=3.2$\times$10$^{-4}$). However, this explanation does not work for NGC~3256-S since the CO emission from the outflow is optically thick.

The lower $\alpha_{\rm CO}$ of ULIRGs (0.8\,$M_\odot\,{\rm (K\,km\,s^{-1}\,pc^2})^{-1}$; \citealt{Downes1998}) is explained because the molecular clouds of these interacting\slash merging systems are not virialized and the width of the line is affected by large scale motions and turbulence.
This argument fits well with the conditions in outflows where the clouds are not self-gravitating (e.g., \citealt{Pereira2020}). This increased velocity dispersion of the clouds increases the observed flux for optically thick lines (as in this outflow) and this results in a reduced $\alpha_{\rm CO}$ factor for the outflow.

We derived an $\alpha_{\rm CO}$ factor which depends on the \hbox{$[$CO\slash H$_2$$]$} abundance ratio. 
The CO abundance in molecular outflows has not been firmly established yet, so this is a major uncertainty for these results. We assumed a standard CO abundance of 3.2$\times$10$^{-4}$ (e.g., \citealt{Bolatto2013}), but CO might be dissociated in molecular outflows and become \ion{C}{i} \citep{Cicone2018, Saito2022}. If so, our $\alpha_{\rm CO}$ values would increase.
However, we note that it is not possible to accurately quantify the CO to \ion{C}{i} dissociation in outflows. The $[$\ion{C}{i}$]$ emission is optically thin (e.g., \citealt{Papadopoulos2004CI}), so the total molecular mass, including molecular gas without CO, derived from its emission depends on the assumed \hbox{$[$\ion{C}{i}\slash H$_2$$]$} abundance ratio which is also uncertain for outflows.

\section{Summary and conclusions}\label{s:summary}

We studied the spatially resolved $^{12}$CO $v$=1--0 and $^{13}$CO \hbox{$v$=1--0} bands, at 4.67\micron\ and 4.78\micron, respectively, around the dust-embedded AGN in the southern nucleus of NGC~3256. 
We modeled the ratios between the P($J$+2) and R($J$) transitions of the CO $v$=1--0 emission using a non-LTE approximation. Using these models, we constrained the column density, kinetic temperature, and $\alpha_{\rm CO}$ conversion factor of the cold molecular gas in the outflow. The main results of the paper are the following:

\begin{enumerate}

\item \textit{CO band classification.} Based on the detection in absorption or emission of the P- and R-branches of the CO $v$=1--0 band, we classify the observed spectra into three categories: (a) both branches in absorption seen basically toward the mid-IR bright nucleus; (b) P-R asymmetry (i.e., P-branch in emission and R-branch in absorption) which is detected along the disk of the galaxy; and (c) both branches in emission detected above and below the disk in the outflow region. We focused on the later to determine the physical conditions of the cold molecular gas in this AGN launched outflow.

\item \textit{Origin of the CO $v$=1--0 band emission.} Detecting the band in emission can be explained as follows: the radiation from the deeply embedded AGN is reprocessed by dust which emits intense mid-IR $\sim$4.7\micron\ continuum radiation. Then, this continuum illuminates the molecular clouds in the outflow and excites the vibrational levels of CO resulting in the $v$=1--0 bands seen in emission in our line of sight.

\item \textit{Temperature gradient in the outflow.} From the spatially resolved map of the $^{12}$CO $v$=1--0 P(2)\slash R(0) ratio, we identify a temperature gradient in the outflow from $>$40\,K in the central 100\,pc to $<$15\,K at 250\,pc. This temperature gradient is similar to that found in the molecular outflow of the local LIRG ESO~320-G030 \citep{Pereira2020}.

\item \textit{Reduced $\alpha_{\rm CO}$ in the outflow.} We derived the $\alpha_{\rm CO}$ conversion factor in the outflow by fitting the observed P($J$+2)\slash R($J$) ratios in eight 100\,pc (0\farcs5) apertures in the outflow region. We used three different approaches (single non-LTE component, two non-LTE components, and $^{13}$CO models).
We find $\alpha_{\rm CO}$ between (0.26--1.09)$\times \frac{3.2 \times10^{-4}}{[{\rm CO}\slash {\rm H_2}]}$ \,$M_\odot$ (K km\,s$^{-1}$\,pc$^2$)$^{-1}$ for the non-LTE modeling with median values of 0.40 $\pm$ 0.15 and 0.61 $\pm$ 0.17 for the single and two components models, respectively. These are 1.3--2 times lower than the ULIRG-like $\alpha_{\rm CO}$ typically assumed for outflows. From the $^{13}$CO models, we find a median $\alpha_{\rm CO}$=(0.19--0.36)$\times\frac{3.2 \times10^{-4}}{[{\rm CO}\slash {\rm H_2}]}\times\frac{[{\rm ^{12}CO}\slash {\rm ^{13}CO}]}{60}$\,$M_\odot$ (K km\,s$^{-1}$\,pc$^2$)$^{-1}$, which is lower than those derived from the non-LTE models, although it is subject to the uncertain $[{\rm ^{12}CO}\slash {\rm ^{13}CO}]$ abundance ratio in the outflow. The most likely explanation for this reduced $\alpha_{\rm CO}$ is that the molecular clouds in the outflow are not virialized.

\item \textit{D$_n$-PAH detection.} We detected a broad ($\sigma$=0.0091 $\pm$ 0.0003\micron) spectral feature at 4.645 $\pm$ 0.002\micron\ which is produced by the aliphatic C-D stretch mode in deuterated PAH (D$_n$-PAH). The D$_n$-PAH emission is detected in the disk of the galaxy and also in the apertures selected in the outflow, although the latter could be background or foreground emission not directly connected to the outflow.

\item \textit{Optically thick $^{12}$CO emission in the outflow.} The observed ratios indicate that the $^{12}$CO $v$=1--0 emission is optically thick, at least for $J$<4, while the $^{13}$CO $v$=1--0 emission remains optically thin.

\end{enumerate}

\begin{acknowledgements}
We thank the referee for their useful comments and suggestions.
We thank A. Alonso-Herrero for the careful reading of the manuscript and useful discussion.
The authors acknowledge the GOALS DD-ERS team for developing their observing program.

MPS acknowledges funding support from the Ram\'on y Cajal programme of the Spanish Ministerio de Ciencia e Innovaci\'on (RYC2021-033094-I).
EG-A thanks the Spanish MICINN for support under projects PID2019-105552RB-C41 and PID2022-137779OB-C41.
IGB acknowledges support from STFC through grant ST/S000488/1.

This work is based on observations made with the NASA/ESA/CSA James Webb Space Telescope. The data were obtained from the Mikulski Archive for Space Telescopes at the Space Telescope Science Institute, which is operated by the Association of Universities for Research in Astronomy, Inc., under NASA contract NAS 5-03127 for JWST; and from the European JWST archive (eJWST) operated by the ESAC Science Data Centre (ESDC) of the European Space Agency. These observations are associated with program \#1328.

This paper makes use of the following ALMA data: ADS/JAO.ALMA\#2015.1.00412.S, ADS/JAO.ALMA\#2015.1.00714.S, ADS/JAO.ALMA\#2018.1.00223.S.
ALMA is a partnership of ESO (representing its member states), NSF (USA) and NINS (Japan), together with NRC (Canada) and NSC and ASIAA (Taiwan) and KASI (Republic of Korea), in cooperation with the Republic of Chile. The Joint ALMA Observatory is operated by ESO, AUI/NRAO and NAOJ.
The National Radio Astronomy Observatory is a facility of the National Science Foundation operated under cooperative agreement by Associated Universities, Inc.

\end{acknowledgements}

\end{document}